\pdfoutput=1
\documentclass [10pt,letterpaper]{JHEP3}
\usepackage{amsmath}
\usepackage{graphicx}
\usepackage[all]{xy}
\usepackage{bm}         			
\usepackage{verbatim}   			

\advance\parskip 1.0pt plus 1.0pt minus 2.0pt   
\def\coeff#1#2{{\textstyle {\frac {#1}{#2}}}}

\def\half{\coeff 12}

\def\N{{\cal N}}

\def\None{\N\,{=}\,1}
\def\Nfour{\N\,{=}\,4}
\def\R{{\mathbb R}}

\def\tr{{\rm tr}}

\def\Nc{N_{\rm c}}
\def\nf{n_{\rm f}}

\def\Z{{\mathbb Z}}

\def\Dslash{{\rlap{\raise 1pt \hbox{$\>/$}}D}}

\def\SN{\mathcal S_{N}}

\title
    {%
    \boldmath
    Large-$N$ volume independence in
    conformal and confining gauge theories
    }%

\author
    {{
    \def\href#1#2{#2}		
     Mithat \"Unsal$^1$\footnote{\email{unsal@slac.stanford.edu}}~
    and Laurence G.~Yaffe$^2$\footnote{\email{yaffe@phys.washington.edu}}
    \\${}^1$SLAC and Physics Department, Stanford University, Stanford, CA 94305
    \\${}^2$Department of Physics, University of Washington, Seattle,
    WA 98195--1560
    }}%

\abstract
    {{ \small
    Consequences of large $N$ volume independence are examined
    in conformal and confining gauge theories.
    In the large $N$ limit,
    gauge theories compactified on $\R^{d-k} \times (S^1)^k$
    are independent of the $S^1$ radii,
    provided the theory has unbroken center symmetry.
    In particular, this implies that a large $N$ gauge theory which,
    on $\R^d$, flows to an IR fixed point,
    retains the infinite correlation length and other scale invariant
    properties of the decompactified theory even when compactified on
    $\R^{d-k} \times (S^1)^k$.
    In other words, finite volume effects are $1/N$ suppressed.
    In lattice formulations of vector-like theories,
    this implies that numerical studies to determine the boundary
    between confined and conformal phases may be performed
    on one-site lattice models.
    In $\Nfour$ supersymmetric Yang-Mills theory,
    the center symmetry realization is a matter of choice:
    the theory on $\R^{4-k}\times (S^1)^k$ has a moduli space
    which contains points with all possible realizations of center symmetry.
    Large $N$ QCD with massive adjoint fermions
    and one or two compactified dimensions
    has a rich phase structure
    with an infinite number of phase transitions
    coalescing in the zero radius limit.
    }}%

\preprint{SLAC-PUB-14160}
\keywords{$1/N$ expansion, lattice QCD, nonperturbative effects}
\begin{document}

\section{Introduction}

Wide classes of large $N$ gauge theories,
when studied on toroidal compactifications of $\R^d$,
have properties which are independent of the compactification radii.
For brevity, we will refer to independence on compactification radii
as \emph{volume independence}.%
\footnote
    {
    Our discussion applies to compactifications on $\R^{d-k} \times (S^1)^k$
    where $k$, the number of compactified dimensions, may range from 1 to $d$.
    If $k < d$, so that some dimensions remain uncompactified,
    then it is an abuse of language to refer to independence
    on the compactification radii as volume independence.
    We hope that our use of this name, which originated in discussions
    of compactifications of $\R^d$ to $(S^1)^d$ with cubic symmetry,
    does not cause confusion.
    In lattice gauge theories, large $N$ equivalence between the
    decompactified theory and a single site  model is sometimes referred to
    as ``large $N$ reduction'' or ``Eguchi-Kawai (EK) reduction''.
    }
Examples (and counterexamples) of
large $N$ volume independence have been discussed since the 1980's
\cite{Eguchi-Kawai,LGY-largeN,BHN,Gonzalez-Arroyo:1982hz,Gonzalez-Arroyo:1982ub},
but there has been a recent resurgence of interest in the subject.

In $SU(N)$ gauge theories on $\R^{d-k} \times (S^1)^k$,
with only adjoint representation matter fields,
large $N$ volume independence holds provided two symmetry realization
conditions are satisfied as $N \to \infty$ \cite{Kovtun:2007py}:
\begin{itemize}
\item
    Translation symmetry is not spontaneously broken.
    \hfill(1a)
\item
    The $(\Z_N)^k$ center symmetry is not spontaneously broken.%
    \footnote
	{%
	Center symmetry transformations are gauge transformations
	which are periodic only up to an element of the center of
	the gauge group.
	For $SU(N)$ theories on $\R^{d-k}\times (S^1)^k$,
	the group of such transformations, modulo gauge transformations
	continuously connected to the identity, is $(\Z_N)^k$.
	Center transformations associated with a particular toroidal cycle
	multiply Wilson loops by a phase factor $z^n$ where $z \in \Z_N$
	and $n$ is the winding number of the loop around the cycle.
	Hence, topologically non-trivial Wilson loops serve as
	order parameters for center symmetry.
	$\Z_N$ center transformations are symmetries of 
	$SU(N)$ gauge theories provided all matter
	fields are in representations (such as the adjoint) with
	vanishing $N$-ality.
	To simplify the presentation, we limit our discussion
	to this class of theories.
	However, it should be noted that there are additional large $N$
	equivalences which relate $SO(N)$, $Sp(N)$ and $SU(N)$
	gauge theories, and which relate theories with matter fields
	in rank-2 symmetric, antisymmetric and bifundamental representations
	to theories with adjoint representation matter
	\cite{Lovelace,KUY1,UY,ASV}.
	}
    \hfill(1b)
\end{itemize}
\addtocounter{equation}{1}
Volume independence applies to the leading large $N$ behavior of
expectation values and connected correlators of topologically
trivial Wilson loops and similar single-trace observables.
See Ref.~\cite{Kovtun:2007py} for more detail.

In $SU(N)$ Yang-Mills theory,
large $N$ volume independence holds as long as all compactification radii
are larger than a critical radius $L_{\rm c} \sim \Lambda^{-1}$ \cite{KNN}.
Volume independence fails below this critical radius
due to a center-symmetry breaking phase transition.
In the case of a single compactified dimension, this is the usual
confinement/deconfinement thermal phase transition.
Two approaches for modifying a lattice gauge theory in order to
suppress this center-symmetry breaking phase transition,
and restore the validity of volume independence down to arbitrarily
small radius (or down to a single site in a lattice regulated theory),
are known.%
\footnote
    {%
    Earlier schemes known as quenched Eguchi-Kawai reduction \cite{BHN}
    and twisted Eguchi-Kawai
    \cite{Gonzalez-Arroyo:1982hz, Gonzalez-Arroyo:1982ub}
    have recently been shown to fail due to nonperturbative effects
    \cite{Bietenholz:2006cz,Teper:2006sp, Azeyanagi:2007su, Bringoltz:2008av}.
    However, see Ref.~\cite{GonzalezArroyo:2010ss} for a recent proposed
    fix for twisted reduction.
    }
One may add double trace deformations to the action involving
absolute squares of Wilson loops wrapping the compactified
directions \cite{Unsal:2008ch}.
Such deformations can suppress spontaneous breaking of center symmetry
while leaving the large $N$ dynamics in the center-symmetric sector
of the theory completely unaffected.
Alternatively, one can add one or more massless adjoint representation fermions,
with periodic (not antiperiodic) boundary conditions
\cite{Kovtun:2007py}.
The fermions modify the effective potential for topologically
non-trivial Wilson loops in a manner which prevents the
breaking of center symmetry even for arbitrarily
small compactification radii.

In this work we discuss the physical basis for large $N$ volume
independence and examine its implications in both
confining and conformal gauge theories.
We highlight the presence of
nonuniformities when the gauge group rank $N \to \infty$
and compactification size $L \to 0$.
For simplicity of presentation, much of our discussion will focus
on the case of a single compactified dimension.
If center symmetry is unbroken
on $\R^{d-1} \times S^1$,
we show that the long distance physics
for finite values of $N$
is sensitive to $L$ only via the combination $NL$.
Specific consequences we discuss include the following:
\begin{enumerate}\advance\itemsep -4pt
\item
    For asymptotically free theories with a strong scale $\Lambda$,
    the compactified theory differs negligibly from the
    decompactified theory provided $NL\Lambda \gg 1$.
    Conversely,
    a dimensionally reduced long distance effective theory
    characterizes the long distance dynamics only when $NL\Lambda \ll 1$.

\item
    For theories which, on $\R^d$, have scale-invariant long distance dynamics,
    compactification on $\R^{d-1}\times S^1$ leads to a correlation length
    of order $NL$; a factor of $N$ times longer than would be naively expected.
    Maximally supersymmetric Yang-Mills theory ($\Nfour$ SYM)
    is a special case where
    the correlation length of the compactified theory
    remains infinite and all realizations of center symmetry coexist.

\item
    For QCD with $\nf$ massless adjoint fermions, the lower boundary of the
    ``conformal window''
    (\emph{i.e.}, the minimal number of flavors $\nf^*$ for which the theory
    flows to a non-trivial IR fixed point) may,
    in the large $N$ limit, be determined
    by studying the compactified theory with arbitrarily small radius.

\item
    For QCD with massive adjoint fermions,
    the symmetry realization and phase structure of the theory,
    compactified on $\R^3 \times S^1$, sensitively depends on the value of
    $NLm$.

\end{enumerate}
In a final section, we discuss the situation with multiple
compactified dimensions.
There are several new issues in this case, but most of the 
$\R^{d-1} \times S^1$ analysis generalizes in a straightforward fashion.

\section{Compactification and center symmetry}\label{scales}

Consider an asymptotically free $SU(N)$ gauge theory
on $\R^3 \times S^1$.
Let $\Lambda$ denote the strong scale of the theory,
and let $\Omega$ denote the holonomy of the gauge field
around the compactified direction.
[In other words, $\Omega(\bm x)$ is the path-ordered exponential
of the line integral of the gauge field around the $S^1$ at
spatial position $\bm x$.]
We assume that the theory has a $\Z_N$ center symmetry,
so our discussion applies to QCD with adjoint representation
fermions [denoted QCD(adj)],%
\footnote
    {
    The results of this discussion also apply to QCD-like
    theories with fermions in symmetric or antisymmetric 
    rank-two tensor representations, or ordinary QCD with
    $\nf$ fundamental representation fermions provided
    $\nf$ is held fixed as $N \to \infty$.
    As noted in Ref.~\cite{Kovtun:2007py} 
    and discussed more fully in Ref.~\cite{Armoni:2007kd},
    an ``emergent'' $\Z_N$ center symmetry appears in 
    the large $N$ limit of theories with symmetric or
    antisymmetric tensor representation matter fields.
    For these theories, the addition of explicit
    double-trace center-symmetry stabilizing terms is needed
    to prevent center symmetry breaking at small $L$
    \cite{Unsal:2008ch}.
    }
provided the number of flavors $\nf$
is below the asymptotic freedom limit, $\nf < \nf^{\rm AF} \equiv 5.5$.
A center symmetry transformation multiplies the holonomy
by a phase factor, $\Omega \to z \, \Omega$ with $z \in \Z_N$.
Hence, Wilson line expectations $\langle \tr\, \Omega^k \rangle$,
for any non-zero $k \bmod N$, are order parameters
for the realization of center symmetry.

\begin{FIGURE}[ht]
    {
    \parbox[c]{\textwidth}
        {
        \begin{center}
        \includegraphics[width=0.8\textwidth]{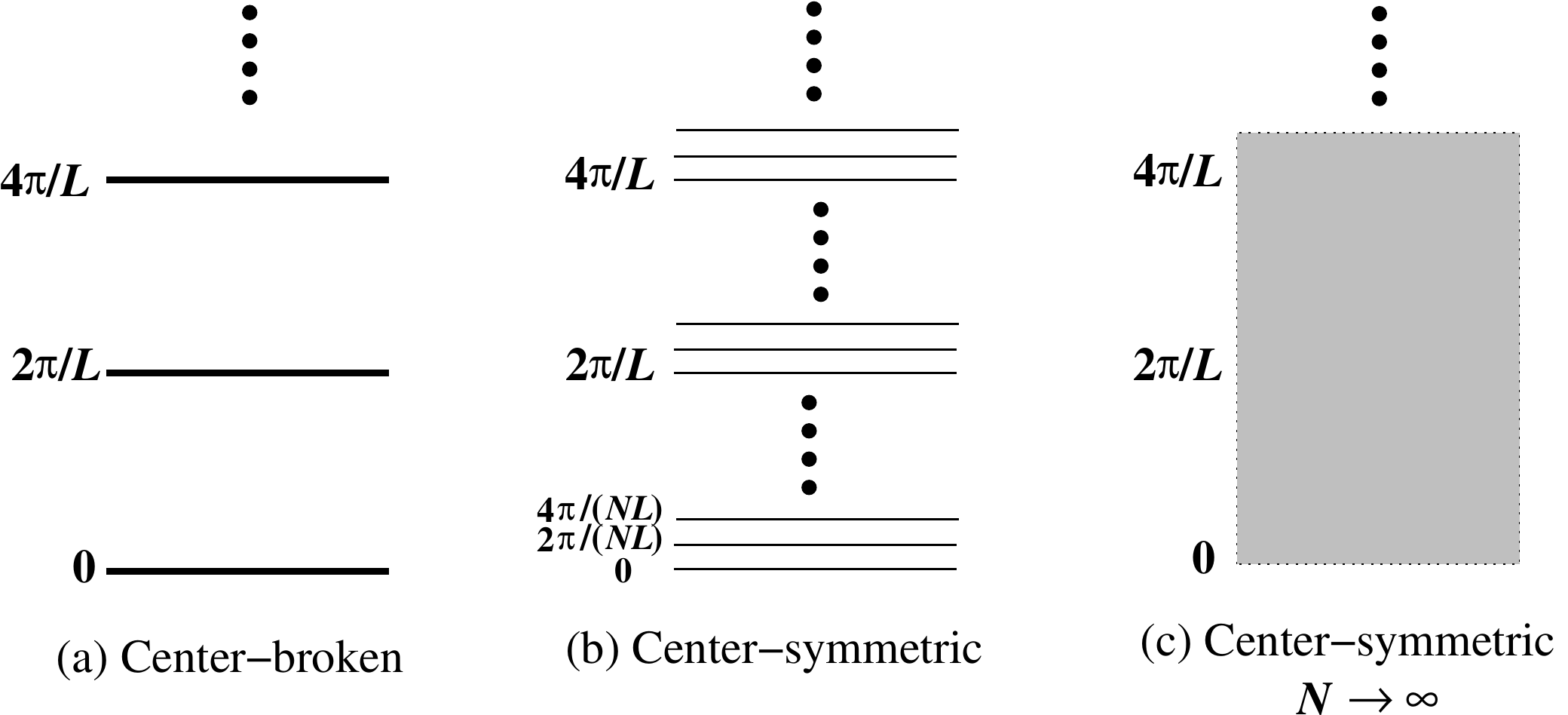}
	\vspace*{-14pt}
        \caption
	    {%
	    Dependence of the Kaluza-Klein spectrum on the
	    realization of center symmetry.
	    (a) Maximally broken center symmetry with
	    holonomy $\Omega = 1$ gives the usual $2\pi/L$ level spacing.
	    (b) Unbroken center symmetry
	    produces a finer $2\pi/(NL)$ level spacing.
	    (c) Sending $N \to \infty$ with unbroken center symmetry
	    leads to continuous spectra.
	    \label {fig:states}
	    }
	\end{center}
	}
    }
\end{FIGURE}

Compactification produces Kaluza-Klein (KK) towers ---
discrete frequency spectra of field modes.
The form of the KK spectra is critically dependent on the
realization of center symmetry, as illustrated in Figure~\ref{fig:states}.
Fig.~\ref{fig:states}a shows the ``standard'' form of Kaluza-Klein towers,
in which allowed momenta are spaced at integer (or half-integer)
multiples of $2\pi/L$.
For adjoint representation fields, each level has an $O(N^2)$ degeneracy.
This is the situation when the holonomy $\Omega = 1$ and the
center symmetry is completely broken.
For physics sensitive to energies small compared to the inverse
compactification radius,
\begin{equation}
    E \ll \frac{1}{L} \,,
\end{equation}
or length scales large compared to $L$,
the non-zero frequency components of all fields may be integrated out,
leading to a dimensionally reduced $3d$ effective theory describing
long distance dynamics \cite{Gross:1980br,Braaten:1995cm,Braaten:1995jr}.

In contrast,
when the center symmetry is unbroken the
eigenvalues of $\Omega$ are (on average)
evenly distributed around the unit circle.
This follows from the vanishing of traces of $\Omega^k$
for all non-zero $k \bmod N$.
Such a non-degenerate eigenvalue distribution of $\Omega$
produces a finer KK spectrum with
spacing of $2\pi/(NL)$ and $O(N)$ degeneracies,
as illustrated in Fig.~\ref{fig:states}b.
To see this, recall that a non-trivial holonomy shifts
the phase acquired by an excitation propagating around the $S^1$,
or equivalently shifts the frequency moding of field components.
When the eigenvalues of $\Omega$ are non-degenerate and uniformly
distributed, every field breaks up into $N$ pieces
with different offsets in the frequency quantization.
The range of energies for which a dimensionally reduced
effective description is valid is now
\begin{equation}
    E \ll \frac{1}{NL} \,.
\label{eq:domain}
\end{equation}
The resulting long distance effective theory,
relevant only for distances large compared to $NL$,
corresponds perturbatively to a $SU(N) \to U(1)^N$
Higgsing of the theory.
The holonomy $\Omega$ acts as an effective $3d$ adjoint representation
Higgs field
which gives
masses in the $[2\pi/NL, 2\pi/L)$ range  to all off-diagonal field components.
In addition to perturbative fluctuations, there are also
non-perturbative topological defects, both self-dual monopoles
and non-self-dual magnetic bions (or monopole-anti\-monopole bound states)
of various charges \cite{Unsal:2007jx}.
A semiclassical analysis (generalizing Polyakov's classic treatment 
\cite {Polyakov:1976fu} to  gauge theory on $\R^3 \times S^1$)
shows that these topological defects,
even when arbitrarily dilute,
can generate a mass gap and area law behavior for spatial Wilson loops,
as discussed in detail for QCD(adj) in
Ref.~\cite{Unsal:2007jx}.

A key point is that integrating out the non-zero frequency modes
perturbatively,
and analyzing the monopole/bion dynamics using semiclassical methods,
is only valid
when the theory is weakly coupled on the scale of $NL$.
In other words, the dimensionally reduced description of the
long distance dynamics,
with semi-classical Abelian confinement, is only valid when
$ N L \Lambda \ll 1 $.

As $N \to \infty$,
with unbroken center symmetry,
the frequency spacing in Kaluza-Klein towers approaches zero and
the spectrum approaches the continuous frequency spectrum
of the decompactified theory,
as illustrated in Fig.~\ref{fig:states}c.
The domain of validity (\ref{eq:domain}) of the $3d$ long
distance description shrinks to zero energy
(and diverging distances).
Observables probing any fixed energy scale become unable to
resolve the vanishing discreteness in the frequency spectrum.
Compactification effects
are $1/N$ suppressed and
the leading large $N$ behavior of expectations,
or connected correlators, of 
single trace, topologically trivial observables
is volume independent.
For dynamics on the scale $\Lambda$,
the effect of compactification is negligible when
$N L \Lambda \gg 1$.
In this regime, there is no weakly coupled description
of the long distance physics.%

Recognition of the connection between unbroken center symmetry
and a $1/N$ suppressed spacing in the KK spectrum does not constitute
a proof of large $N$ volume independence.
For that, one must compare the large $N$ loop equations in
lattice regularized theories \cite{KUY1},
or the $N = \infty$ classical dynamics
generated by appropriate coherent states \cite{KUY2}.
But consideration of the KK spectrum does provide a
simple physical understanding of the origin of large $N$
volume independence, and clarifies the relevant scales
which control the approach to the $N = \infty$ limit.

To reiterate, for theories on $\R^3 \times S^1$,
when the center symmetry is not spontaneously broken
the physically relevant length scale appearing in finite volume
effects is not $L$, but rather $NL$.
For confining theories with a strong scale $\Lambda$, there are two
distinct characteristic regimes:
\begin{align}
    NL\Lambda \ll 1\,, & \qquad
    \mbox{semi-classical, Abelian confinement  $\Longrightarrow$ volume dependence;}
\\
    NL\Lambda \gg  1\,, & \qquad
    \mbox{non-Abelian confinement   $\Longrightarrow$ volume independence.}
\label{eq:ana}
\end{align}

\section {Conformal theories and conformal windows}\label{sec:window}

Interest in possible extensions to the standard model involving
strongly coupled, nearly conformal sectors
\cite{Holdom:1981rm,
Yamawaki:1985zg,
Appelquist:1986an,
Hill:2002ap,
Sannino:2004qp,
Luty:2004ye,
Sannino:2009za} 
has stimulated substantial efforts to determine ``conformal windows''
in QCD-like gauge theories with varying fermion content 
\cite{Appelquist:2007hu,
Deuzeman:2009mh,
Fodor:2008hn,
DeGrand:2008kx,
Catterall:2007yx,
Catterall:2008qk,
Hietanen:2008mr,
DelDebbio:2008zf,
DelDebbio:2009fd,
Hietanen:2009az}.
Numerical simulations must, of course, work with UV and IR regulated
theories, and finite volume toroidal compactifications are virtually
always used as the IR regular of choice.
As with all numerical simulations of lattice gauge theories,
efforts to find the boundary between confining and conformal behavior
must adequately control multiple sources of systematic error:
finite volume effects, non-zero lattice spacing artifacts,
and extrapolations to the chiral limit.

For theories with infinite correlation length
(when defined on $\R^d$),
compactification of a spatial dimension introduces a new
length scale, the compactification size $L$.
This modifies the behavior of fluctuations with wavelengths
comparable or larger than $L$ and typically produces a
finite correlation length of order $L$.
But for theories satisfying large $N$ volume independence,
for the reasons discussed in the previous section,
the characteristic size of a compactification-induced
correlation length is not the physical size $L$,
but instead equals $L$ times a positive power of $N$.
This means that finite volume effects are $1/N$ suppressed
as long as the center symmetry is unbroken.
Even for simulations at modest values of $N$, this additional suppression
of finite volume effects decreases the lattice size required to
achieve acceptably small systematic errors.
For larger values of $N$, it implies that very small lattices will be
sufficient to extract infinite volume observables.
The basic point we wish to stress is that large $N$ volume independence 
is not restricted to confining phases;
it is equally applicable to conformal phases provided
the symmetry realization conditions (1) are satisfied.

\begin{FIGURE}[ht]
    {
    \parbox[c]{\textwidth}
        {
        \begin{center}
        \includegraphics[scale=0.55]{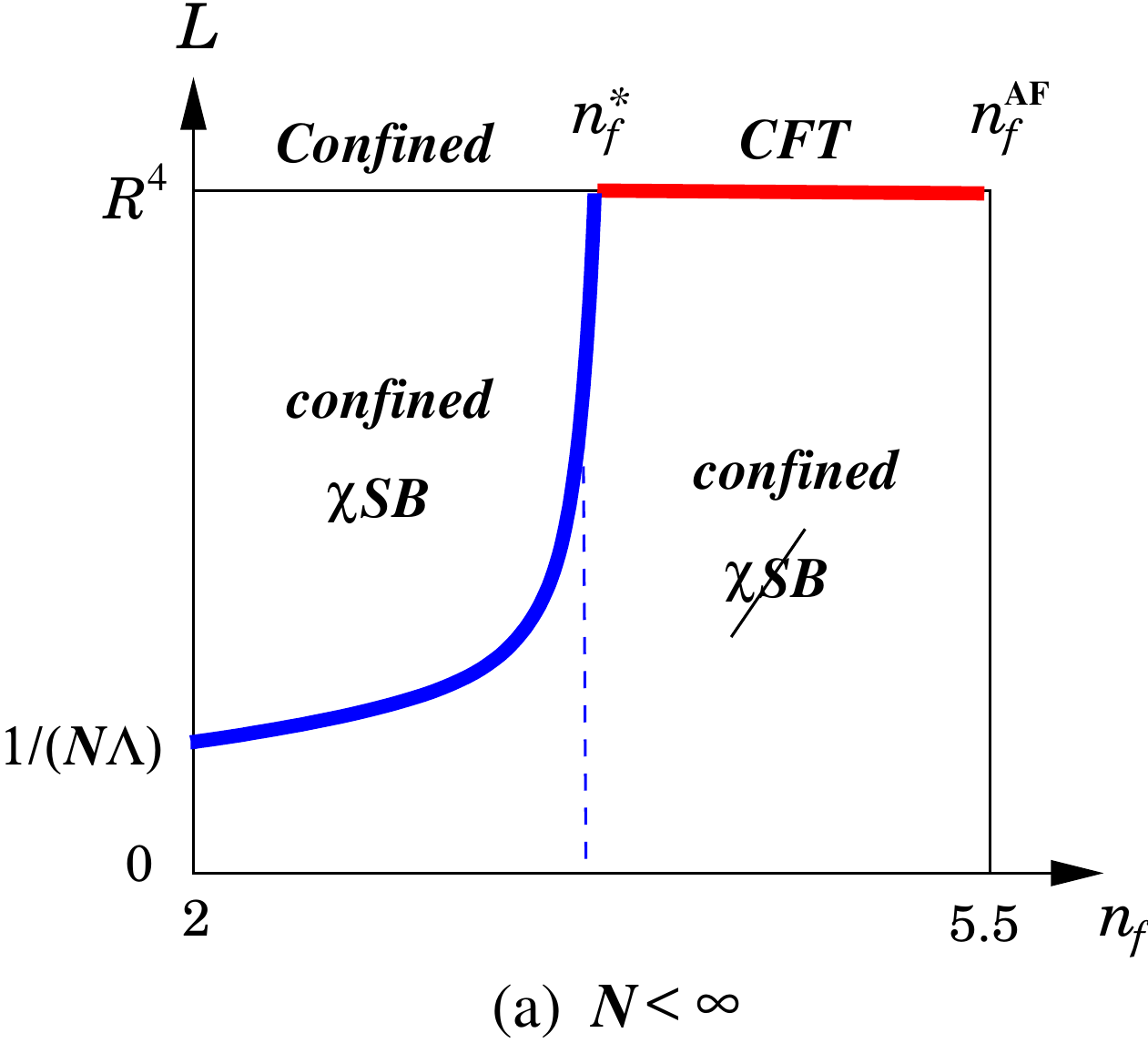}
	\hfil
	\hfil
        \includegraphics[scale=0.55]{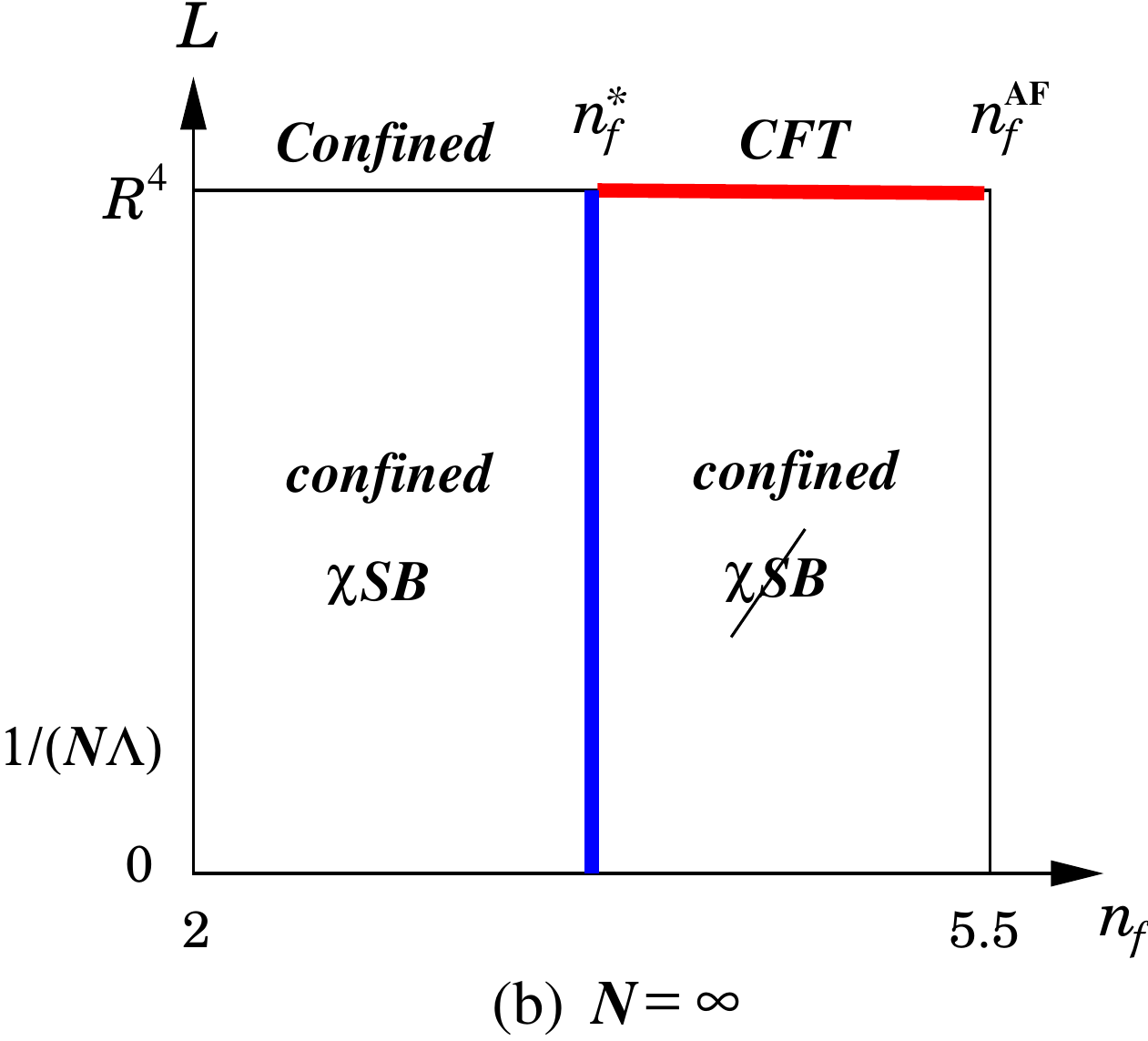}
	\vspace*{-15pt}
        \end{center}
        \caption
	    {
	    Contrasting finite $N$ and infinite $N$
	    phase diagrams of massless QCD(adj) on $\R^3 \times S^1$,
	    as a function of the compactification size $L$ and the
	    number of fermion flavors $\nf$.
	    In the decompactified limit,
	    one expects a confining phase with spontaneously broken
	    chiral symmetry for sufficiently small $\nf$,
	    and a conformal phase with unbroken chiral symmetry
	    in the window $\nf^* \le \nf < \nf^{\rm AF}$
	    (with $\nf^{\rm AF}$ the asymptotic freedom limit).
	    For finite $N$ (left), as one decreases $L$
	    the chiral transition line must bend
	    and approach an intercept at an
	    ``unconventional'' scale of $1/(N\Lambda)$.
	    To the right of this line is a phase with
	    unbroken continuous  chiral symmetry and finite,
	    compactification-induced correlation length,
	    smoothly connecting the analytically
	    tractable $NL\Lambda \ll 1$ region to the
	    conformal portion of the $L=\infty$ boundary.
	    At $N=\infty$ (right), the theory exhibits
	    volume independence in both the chirally broken
	    and chirally symmetric phases.
	    The phase transition line extends straight down
	    from $L = \infty$ and $\nf = \nf^*$.
	    This implies that numerical studies on very small lattices
	    can be used to determine
	    the conformal window boundary $\nf^*$.
	    }
	\label {fig:phase}
        }
    }
\end{FIGURE}

As an example, consider QCD(adj) with massless fermions,
and one dimension periodically compactified,
as a function of the number of fermions $\nf$
and compactification size $L$.
Figure \ref{fig:phase} illustrates the situation.%
\footnote
    {
    This figures treats the number of fermion flavors
    $\nf$ as a continuous variable.
    In the Euclidean theory,
    one can define non-integer $\nf$ by taking
    the fermion determinant,
    which is real and positive in QCD(adj),
    to a fractional power.
    }
In the decompactified limit, $L = \infty$,
one expects a confining phase with spontaneously broken
chiral symmetry for sufficiently small $\nf$,
and a conformal phase (\emph{i.e.}, a phase in which the
theory flows to a non-trivial IR fixed point)
with unbroken chiral symmetry
in some window $\nf^* \le \nf < \nf^{\rm AF}$,
where $\nf^{\rm AF}$ is the asymptotic freedom limit.
Consequently,
as $L$ decreases from infinity,
there should be both chirally symmetric and chirally
asymmetric phases extending into the
$(L,\nf)$-plane phase diagram.%
\footnote
    {%
    More precisely, there is a continuous
    non-Abelian chiral symmetry when $\nf > 1$,
    plus a flavor-independent discrete chiral symmetry.
    Our discussion focuses on the continuous chiral symmetry.
    Unlike the non-Abelian chiral symmetry,
    in the compactified theory the discrete chiral symmetry can be
    spontaneously broken even at weak coupling \cite{Poppitz:2009tw}.
    An unconventional order parameter for the discrete chiral symmetry
    is a topological disorder (or monopole) operator.
    }

For finite $N$ and sufficiently small $L$
[small compared to $1/(N\Lambda)$],
one can reliably analyze the theory using perturbative
and semiclassical methods \cite{Unsal:2007jx}, as mentioned earlier.
One finds unbroken continuous chiral symmetry, broken discrete chiral symmetry, 
a non-zero mass gap, and area-law behavior for large topologically trivial
spatial Wilson loops.%

   The simplest, most plausible, scenario is that
the chirally symmetric phase at small $L$ (and any $\nf$)
smoothly connects to the chirally symmetric conformal
phase at $L = \infty$ and $\nf \ge \nf^*$.
The chiral transition line separating these two phases
must bend as $L$ decreases, as shown,
and approach an intercept at an ``unconventional''
scale of $1/(N\Lambda)$ \cite{Unsal:2008eg}.
This is the scale below which the long distance
semiclassical analysis is valid.
To the left of this line, one has a typical
confining phase with spontaneously broken continuous chiral symmetry.
The mass gap (inverse correlation length)
vanishes due to the presence of Goldstone bosons.
The spatial string tension (or area law coefficient
for large spatial Wilson loops) will have finite,
non-vanishing limits as $L \to \infty$.
To the right of this line one has a phase with
unbroken continuous chiral symmetry and finite,
compactification-induced correlation length
which diverges as $L\to\infty$.
Similarly, this phase will
have a non-zero spatial string tension for finite $L$,
which vanishes as $L \to \infty$.

Massless adjoint fermions,
with periodic boundary conditions on $\R^3 \times S^1$,
prevent the spontaneous breaking of center symmetry
\cite{Davies:2000nw, Kovtun:2007py}.
Consequently,
at $N=\infty$ (illustrated in Fig.~\ref{fig:phase}b),
both the chirally broken and chirally symmetric phases
will exhibit volume independence.
The chirally asymmetric phase will have a finite
correlation length and non-zero spatial string tension,
both independent of $L$.
The chirally symmetric phase will have
an infinite correlation length, irrespective of $L$.
In this long-distance conformal phase,
large Wilson loop expectation values will show
Coulombic behavior.
For rectangular $R \times T$ loops,
\begin{equation}
    \lim_{T \rightarrow \infty}
    \frac{-1}{T} \> \ln  \big\langle W(R\times T) \big\rangle
    \sim
    \begin{cases}
	      \sigma \, R \,, & \mbox{confined phase};
	      \\
	      -{g^2_*}/{R} \,,  &   \mbox{IR conformal phase},
    \end{cases}
\end{equation}
with $\sigma$ and $g_*^2$ having large $N$ limits which are
independent of the compactification size $L$.
The phase transition line must extend straight down
from $L = \infty$ and $\nf = \nf^*$.
The value of $\nf^*$ in the decompactified theory can
be extracted from numerical studies on lattices with arbitrarily
small $L$.
In effect, large $N$ volume independence allows one to trade
extrapolations in lattice volume for extrapolations in the number
of colors, $N$.

\section{\boldmath $\N=4$ SYM}

When a $\Z_N$ center-symmetric theory is compactified on $\R^{d-1} \times S^1$,
the dynamics of the theory may generate an effective potential for the Wilson line
holonomy which causes its eigenvalues to attract, as in pure Yang-Mills theory,
leading to spontaneous breaking of center symmetry.
Alternatively, repulsive interactions between eigenvalues may be generated,
as in massless QCD(adj) (with periodic fermion boundary conditions),
or engineered (by the addition of explicit stabilizing terms),
thereby preventing center symmetry breaking.
But there is a third possibility, which is realized by maximally
supersymmetric Yang-Mills ($\Nfour$ SYM) theory:
a strictly vanishing effective potential for the Wilson line holonomy.

Consider $\Nfour$ SYM theory with gauge group $SU(N)$ on $\R^4$.
The theory possess an $SU(4)_R$ {\it R}-symmetry group, and contains
adjoint representation scalars $\Phi_{[IJ]}$ and fermions $\lambda_I$
in the $\bf 6$ and $\bf 4$
of $SU(4)_R$, respectively.
The renormalization group beta function vanishes identically,
so the gauge coupling $g^2$ is a scale independent physical parameter of the theory.
There is a continuous moduli space of vacua,
\begin{equation}
    {\cal M}_{\R^4} = \R^{6N}/\SN \,,
\label{eq:N=4moduli}
\end{equation}
corresponding to mutually commuting scalar field expectation values
with completely arbitrary values for their eigenvalues
(modulo Weyl group permutations).
The physical theory at the origin of moduli space is an
interacting non-Abelian CFT.
Generic points in moduli space correspond to Higgsing of the $SU(N)$
gauge group down to a maximal Abelian $U(1)^{N-1}$ subgroup.

When compactified on $\R^3 \times S^1$
with periodic spin connection
(\emph {i.e.}, periodic boundary conditions for fermions),
the theory has a $\Z_N$ center symmetry and remains supersymmetric.
The behavior of the compactified theory was previously discussed by
Seiberg \cite{Seiberg:1997ax}.
He argued that the theory at the origin of the moduli space
(\ref{eq:N=4moduli})
flows to three-dimensional $\mathcal N\,{=}\,8$ superconformal SYM theory
at low energies, $E \ll g_3^2 \equiv g^2/L$,
with an emergent $SO(8)_R$ symmetry.
The compactified theory is manifestly not volume independent:
results depend on the compactification scale $L$ and long distance properties
on $\R^3 \times S^1$ differ from the uncompactified theory on $\R^4$.

However, the compactified theory on $\R^3 \times S^1$ has a \emph{larger}
moduli space than does the uncompactified theory.
Due to the maximal $\Nfour$ supersymmetry,
no superpotential for the Wilson line holonomy,
either perturbative or non-perturbative, is generated.
Consequently,
compactification adds a new branch to the moduli space,
with the holonomy $\Omega$ behaving as an adjoint Higgs field.
The moduli space of $\N=4$ SYM on ${\R^3 \times S^1}$ is
\begin{equation}
    {\cal M}_{\R^3 \times S^1}=  [\R^{6N} \times ({\widetilde S^1})^N]/\SN \,,
\label{eq:N=4compact}
\end{equation}
where $\widetilde S^1$ is the dual circle on which eigenvalues of the
Wilson line holonomy $\Omega$ reside.%
\footnote
    {%
    More generally, when compactified on
    $ \R^{4-k} \times (S^1)^k $,
    the quantum moduli space of $\Nfour$ SYM theory is
    $
    {\cal M}_{\R^{4-k}\times (S^1)^k}=  [\R^{6} \times ({\widetilde S^1})^k]^N/\SN
    $,
    corresponding to arbitrary eigenvalues for the six
    scalar fields and $k$ independent Wilson line holonomies,
    all mutually commuting, modulo Weyl group permutations.
    }
In this compactified theory, the realization of the $\Z_N$ center symmetry
is entirely a matter of choice.
Generic points in the moduli space (\ref{eq:N=4compact}) will involve
a set of eigenvalues $\{ e^{i \theta_a} \}$, $a = 1,\cdots,N$, of $\Omega$
which are not invariant under any $\Z_N$ transformation
(other than the identity), and hence completely break the center symmetry.
But since the choice of eigenvalues for $\Omega$ is arbitrary,
the moduli space also includes points where the set of eigenvalues is
completely $\Z_N$ symmetric, as well as points where the set of eigenvalues
is invariant only under some subgroup of $\Z_N$ (if $N$ is composite).

Prior discussion \cite{Seiberg:1997ax} of compactification in
$\Nfour$ SYM has focused on the case where $\Omega = 1$.
With all eigenvalues of $\Omega$ clustered at a single point,
volume independence fails, as noted above.
Here, we instead wish to examine the situation at center-symmetric points
in moduli space where
eigenvalues are evenly distributed around the unit circle and%
\footnote
    {
    For $N$ even, this should be multiplied by $e^{i \pi/N}$
    so that $\det\Omega = 1$.
    }
\begin{equation}
    \Omega
    =
    {\rm diag}
    \left(
    1,\, e^{{2 \pi i}/{N}},\, e^{{4 \pi i}/{N}},\, \cdots,\, e^{2\pi i(N-1)/N} 
    \right) .
\label{centersym}
\end{equation}
For simplicity, focus on the case where
scalar field expectations are small compared to $1/L$.
For finite $N$,
at such center symmetric points in the moduli space,
the dynamics Abelianizes at large distances and reduces
to an effective $U(1)^{N-1}$ gauge theory.
This will be a valid description for distances
large compared to the inverse of the (lightest) $W$-boson mass,
\begin{equation}
m_W = \frac{2 \pi}{LN} \,.
\end{equation}
Off-diagonal components of all fields acquire masses of order $m_W$ or greater,
due to their coupling to the component of the gauge field in the compactified
direction.
The surviving components are those aligned along the Cartan subalgebra of $SU(N)$.
The effective theory describing physics at energies well below $m_W$ is just
a free $U(1)^{N-1}$ Abelian gauge theory with neutral massless fermions and scalars,
\begin{equation}
    {\cal L}
    =
    \frac{1}{g_3^2} \sum_{a =1}^{N-1}
    \left[
	\tfrac{1}{4} (F_{ij}^a)^2
	+ \tfrac{1}{2} (\partial_{i} A_{4}^a)^2
	+  \tfrac{1}{2} \sum_{I<J=1}^4 (\partial_{i} \Phi_{IJ}^a)^2
	+ \sum_{I=1}^{4}  i \, \bar \lambda_I^{a} \sigma_i \partial_i \lambda_I^{a}
    \right].
\label{abel2}
\end{equation}
Here $g_3^2 \equiv g^2/L$ and $A_4$ is the component of the
gauge field in the compactified direction.

One may wonder whether non-perturbative effects will generate
mass terms for the photons or the neutral matter fields.
This is precisely what happens in QCD(adj)
due to the effects of topological excitations.
In $\Nfour$ SYM with, for example, gauge group $SU(2)$ Higgsed down to $U(1)$,  
the index theorem dictates that a  BPS  monopole operator has the form
${\cal M}_1 =e^{-|\phi| + i \sigma} \det_{IJ} \lambda^I \lambda^J$ 
where $\sigma$ is a scalar field dual to the $3d$ gauge field
and $e^{\pm i \phi}$ are the eigenvalues of $\Omega$.
However, it is not hard to see that the Yukawa couplings present in
$\Nfour$ SYM, but absent in QCD(adj),
lift the fermion zero modes in such a way that no non-perturbative
superpotential can be generated.
In other words, the non-renormalization theorems associated with the
large amount of supersymmetry guarantee the masslessness
of low energy excitations.
Thus, the action (\ref{abel2}), which describes
the physics of the massless modes depicted in Fig.\,\ref{fig:states}b,
is not only valid in perturbation theory,
it is also the correct non-perturbative description of
long distance physics.

At face value, the above description seems to be at odds with volume independence.
The theory at the origin of the moduli space on $\R^4$ is non-Abelian,
but the finite $N$ center-symmetric theory on  $\R^3 \times S^1$
exhibits dynamical Abelianization.
The resolution of this apparent puzzle reflects the interplay of the
$N\to\infty $ limit and the compact topology of the
${\widetilde S}^1$ portion of moduli space.
Since ${\widetilde S}^1$ is compact, as $N\to \infty$ the $\Z_N$ symmetrically
distributed eigenvalues become arbitrarily dense and form a continuum.
The domain of validity of the effective $3d$ Abelian theory
is restricted to energies small compared to $m_W = 2\pi/(NL)$ and
shrinks to zero as $N \to \infty$.

To summarize, if one considers $\Z_N$ center-symmetric
vacua in $\Nfour$ SYM theory, compactified on $\R^3 \times S^1$,
then the appropriate long distance effective theory is not the $3d$ interacting 
 superconformal theory with enhanced $R$-symmetry discussed by Seiberg
\cite{Seiberg:1997ax}.
With a center-symmetric compactification,
this enhancement of $R$-symmetry does not take place and the original
$SU(4)_R$ symmetry is apparent at all length scales.
In contrast to the QCD-like theories discussed earlier,
$\Nfour$ SYM never develops a finite compactification-induced
correlation length.
Large $N$ volume independence does apply to center-symmetric vacua
in $\Nfour$ SYM theory and implies that the compactified center-symmetric
theory will, at $N = \infty$, exactly reproduce properties of the
uncompactified CFT on $\R^4$.
But, just as in the discussion of confining theories in section \ref{scales},
there is non-uniformity between the $N \to \infty$
and $L \to 0$ limits.
This is reflected in the characteristic regimes:
\begin{align}
    E \ll {2 \pi}/(LN)\,,& \qquad
	\mbox{low energy, Abelian free Coulomb};
\\
    E \gg {2 \pi}/(LN)\,,& \qquad
	\mbox{high energy, non-Abelian CFT},
\label{abel}
\end{align}
in complete accord with Fig.~\ref{fig:states}.%
\footnote
    {%
    In the strong coupling domain where $\lambda \equiv g^2 N \gg 1$,
    an AdS/CFT analysis \cite{Poppitz:2010bt} shows that
    the criterion for the high-energy non-Abelian CFT regime becomes
    $ E \gg  \lambda / (LN)$.
    }

\section{Massive QCD(adj)}

In $SU(N)$ Yang-Mills theory, compactified on $\R^3 \times S^1$
with compactification size $L \lesssim 1/\Lambda$,
the one-loop contribution to the  Wilson line effective potential
produces eigenvalue attraction and spontaneous breaking of center symmetry.
This symmetry realization is stable if very massive adjoint representation
fermions are added to the theory, since their effects
are negligible when the fermion mass $m$ is large compared to $1/L$.
But when massless adjoint representation fermions are present,
with periodic boundary conditions,
their contribution to the Wilson line effective potential
favors eigenvalue repulsion and overcomes the attractive pure gauge
contribution, leading to unbroken center symmetry.
Consequently, it is inevitable that QCD(adj) on $\R^3 \times S^1$,
with periodic spin connection, exhibits one or more phase transitions
as the fermion mass $m$ is varied for fixed small compactification radius $L$.

In asymptotically free QCD(adj),
one would naively expect the behavior of the massive theory
to closely resemble the massless limit when the mass $m$
is small compared to $1/L$.
As we next discuss, this expectation is too simplistic.
Quite a rich phase diagram emerges as the fermion mass 
and the compactification radius are varied.%
\footnote
    {%
    Related analysis of QCD(adj)
    on $\R^3 \times S^1$ has been performed in 
    Refs.~\cite{Myers:2008zm, Meisinger:2009ne, Bringoltz:2009mi}.
    Numerical simulations on a finite lattice mimicking this geometry 
    are reported in Ref.~\cite{Cossu:2009sq}.
    }

When $N L \Lambda \ll 1$, non-zero frequency Kaluza-Klein modes
are weakly coupled and may be integrated out perturbatively.%
\footnote
    {
    When $\Omega \sim 1$ the relevant weak coupling criterion is
    $\Lambda L \ll 1$.
    But, as discussed in section \ref{scales},
    when the center symmetry is unbroken the KK-mode frequency spacing
    is smaller by a factor of $1/N$, so the weak coupling condition
    applies to the length scale $NL$ instead of $L$.
    }
With $\nf$ adjoint fermions having a common mass $m$,
the resulting one-loop potential
may be conveniently expressed in the form
\begin{align}
    V[\Omega]&=
    \frac{2}{\pi^2L^4}
    \sum_{n=1}^{\infty} \>
	\left[ - 1 + \half \nf \, (nLm)^2 K_2(n L m) \right]
	\frac{\left|\tr\, \Omega^n \right|^2}{n^4} \,.
\label{eq:pot1}
\end{align}
Here $K_2(z)$ is the modified Bessel function of the second kind,
with asymptotic behavior
\begin{eqnarray}
    K_2(z) \sim
    \left\{ \begin{array}{ll}
     \frac{2}{z^2} -2 + O(z^2)\,, \quad &  z \ll 1 \,;
     \\
     \sqrt \frac{\pi}{2z} \; e^{-z}\,, \quad & z \gg 1 \,.
    \end{array} \right.
\label{asymptotes}
\end{eqnarray}
As the mass $m \to \infty$, the fermions decouple
and the effective potential (\ref{eq:pot1})
reduces to the pure gauge result,
$
    V_{\rm YM}[\Omega] = 
    -\frac{2}{\pi^2L^4}
    \sum_{n=1}^{\infty}
    {\left|\tr\, \Omega^n \right|^2} /n^4
$,
up to exponentially small $O(e^{-Lm})$ corrections.
In the opposite limit of massless fermions,
the coefficient of the $|\tr\,\Omega^n|^2$ in 
the series (\ref{eq:pot1}) reduces to $[-1 + \nf]/n^4$,
previously found in \cite{Kovtun:2007py},
and
$
    V[\Omega] \to -(\nf-1) \, V_{\rm YM}[\Omega]
$.

The $n$'th term in the series (\ref{eq:pot1})
is evidently an effective mass term for the
winding number $n$ Wilson line $\tr(\Omega^n)$.
For arbitrary $SU(N)$ matrices,
$\tr (\Omega^n)$ is reducible to lower order traces when $|n| > N/2$,
but traces with windings up to $|n| = \lfloor N/2 \rfloor$
may be regarded as independent.
If the effective masses
\begin{equation}
    m_n^2
    \equiv
    \frac{4}{\pi^2 n^4}
    \left[  -1 + \half {\nf} \, z_n^2 K_2(z_n) \right],
    \qquad  z_n \equiv  nLm \,,
\label{eq:m_n}
\end{equation}
are positive for all $n \le \lfloor N/2 \rfloor$,
then the minimum of the one-loop potential lies at the center symmetric point
where $\tr\>\Omega^n = 0$ for all non-zero $n \bmod N$.
If some of these effective masses are negative, then 
the corresponding Wilson lines will develop non-zero expectation values,
implying spontaneous breaking of center symmetry.
Since $z^2 K_2(z)$ decreases monotonically for $z > 0$,
if the first mass $m_1^2$ is negative then so are all higher masses.
In this case, the $\Z_N$ center symmetry is completely broken.
If $N$ is large (and composite), there can be a plethora of
intermediate phases where the center symmetry partially breaks
to different discrete subgroups.
The effective mass squared (\ref{eq:m_n}) vanishes when
$z_n$ exceeds a threshold $z_*$ whose value depends on $\nf$,
\begin{equation}
z_{*} (\nf) = \{ 2.027, \; 2.709, \; 3.154, \; 3.484 \} \,,
\end{equation}
for  $\nf=2, 3, 4, 5$.%
\footnote
    {%
    The single flavor case $\nf =1$ is discussed separately below.
    In QCD(adj),
    asymptotic freedom is lost at $\nf^{\rm AF} = 5.5$.
    }
Consequently,
the Wilson line with winding number $n \le \lfloor N/2 \rfloor$
becomes unstable when the compactification size $L$ exceeds
\begin{equation}
    L_n \equiv  \frac{z_{*}}{n m} \,,
\label{eq:seq}
\end{equation}
Since
$
   L_{\lfloor {N}/{2} \rfloor} < \cdots < L_2 < L_1
$,
if $L$ lies in the interval $[L_k, L_{k-1}]$
then $m_k^2 < 0$ and $m_{k-1}^2 > 0$,
implying that Wilson lines wrapping fewer than $k$ times are stable,
while loops with $k$ up to $\lfloor N/2 \rfloor$ windings
are unstable.
When $ L < L_{\lfloor {N}/{2} \rfloor}$,
all independent Wilson lines are stabilized,
and center symmetry is unbroken.

When $\nf > 1$,
for any finite mass $m$, as one increases
$L$ from zero the initial phase is fully $\Z_N$ symmetric.
For sufficiently small $L$ a perturbative analysis is reliable
for any value of $m$.
One may regard this regime as one in which the $SU(N)$ gauge group
is Higgsed down to the maximal Abelian subgroup $U(1)^{N-1}$.%
\footnote
    {
    At large $m$ and small $L$, confinement is the result
    of  monopoles carrying a net magnetic charge and  topological charge $1/N$.
    At $m=0$ and small $L$, confinement results from magnetic ``bions,''
    magnetically charged, topologically neutral combinations of
    monopoles and antimonopoles of differing types, 
    which become bound due to fermion zero mode exchange \cite{Unsal:2007jx}. 
    Turning on a non-zero fermion mass lifts the fermion zero modes
    and allows the bions to unbind, smoothly converting the small $m$
    bion-induced confinement into monopole-induced confinement at large $m$. 
    }
When $L$ exceeds $L_{\lfloor {N}/{2} \rfloor}$, the highest
(independent) winding mode becomes unstable and develops a
non-zero expectation value --- provided the weak coupling
condition $N L \Lambda \ll 1$ which underlies this analysis
is valid.
When $L \sim L_{\lfloor N/2 \rfloor} = O(1/Nm)$, this 
condition implies that the fermion must be heavy,
$m \gg \Lambda$.
As $L$ continues to increase (with $m \gg \Lambda$),
successively lower winding modes become unstable and
will develop expectation values 
each time $L$ passes a threshold $L_k$.%
\footnote
    {
    The slopes of the phase transition lines (\ref{eq:seq})
    for small $L$ agree with Ref.~\cite{Hollowood:2009sy},
    which studied the same class of theories on $S^3 \times S^1$.
    Our values (\ref{eq:seq}) agree with the  numerical values on
    the $mL$ axis of Fig.~4b of Ref.~\cite{Hollowood:2009sy}.
    }
All modes are locally unstable when $L$ exceeds $L_1$,
so the center symmetry will be completely broken
when $L > L_1$ (and $NL\Lambda \ll 1$).

This analysis is valid in the small-$L$,
large-$m$ corner of the $(L,m)$-plane phase diagram,
when $\nf > 1$.
The behavior in other regions of the phase diagram,
especially the small mass regime,
depends on the number of fermion flavors.
We will discuss separately three cases:
multiple flavors below the conformal window, $2 \le \nf < \nf^*$;
a single flavor, $\nf = 1$;
and multiple flavors within the conformal window,
$\nf^* \le \nf < \nf^{\rm AF}$.

\subsection*{Multiple flavors: \boldmath $2 \le \nf < \nf^*$}

\begin{FIGURE}[ht]
    {
    \parbox[c]{\textwidth}
        {
        \begin{center}
	\vspace*{5pt}
        \includegraphics[scale=0.5]{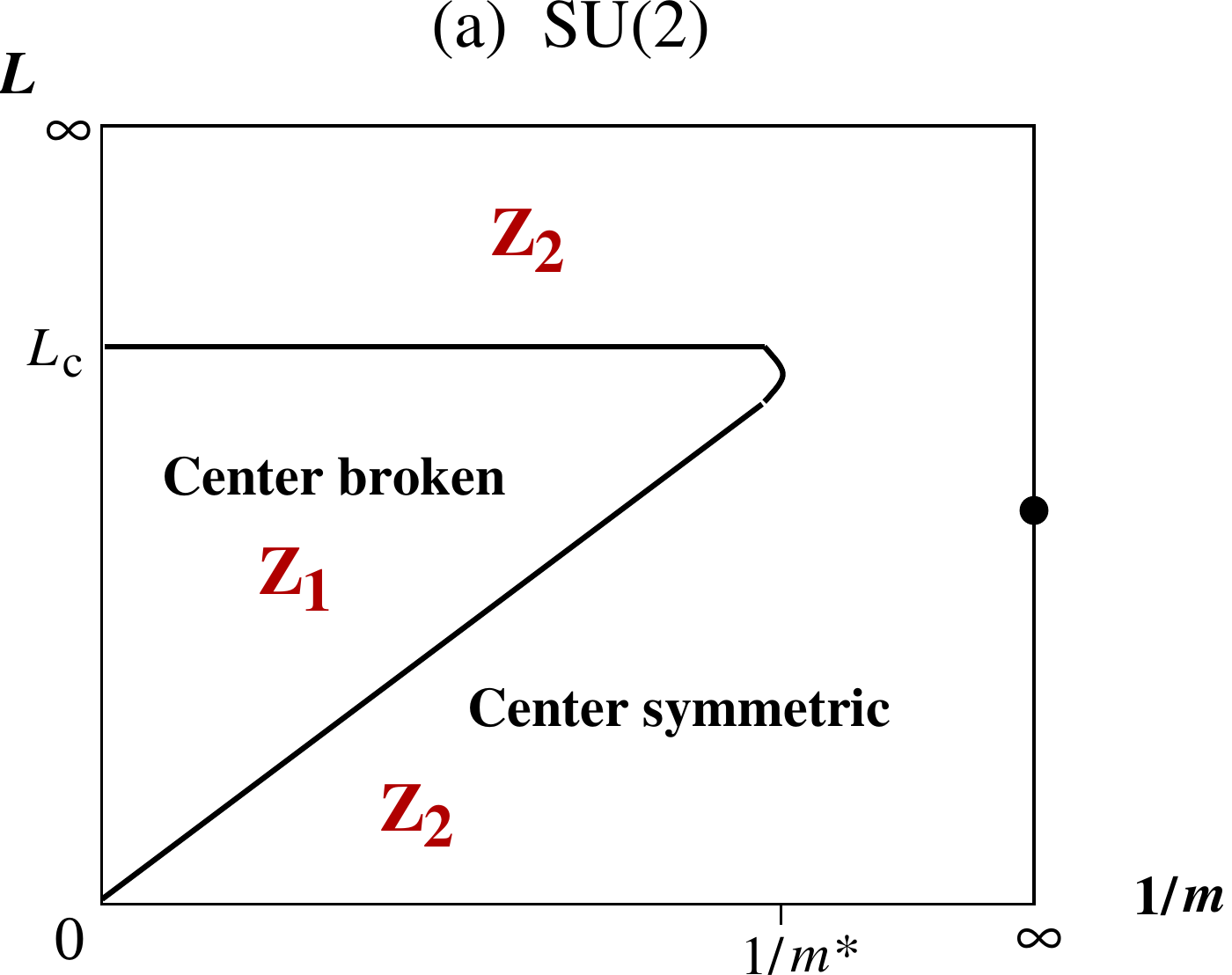}
	\hfil
	\hfil
        \includegraphics[scale=0.5]{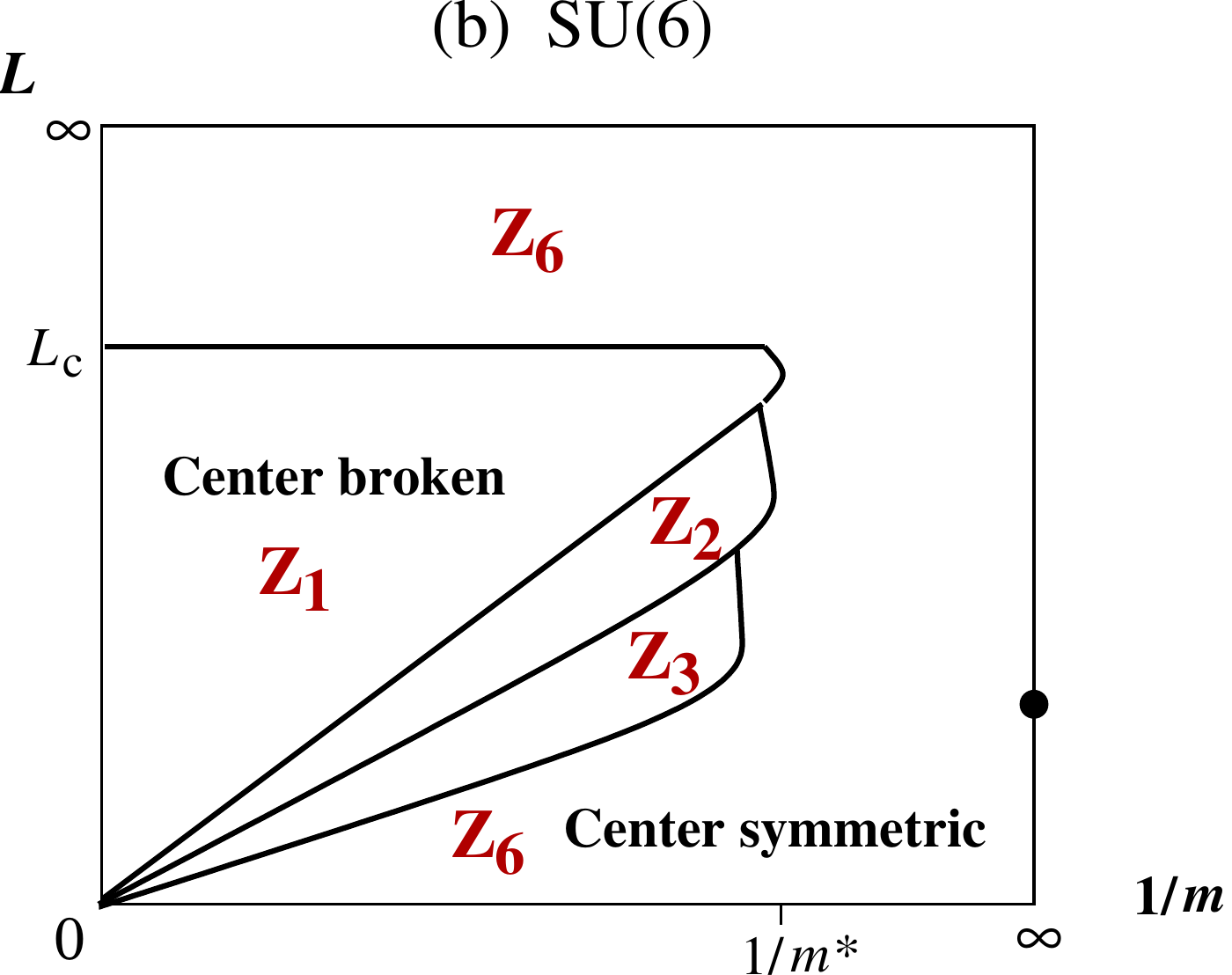}
	\vspace*{-10pt}
        \caption
	{
	Center symmetry realization of QCD(adj) on $\R^3 \times S^1$
	with $2 \le \nf < \nf^*$ and periodic spin connection,
	as a function of compactification size $L$ and inverse
	fermion mass $1/m$.
	Illustrated are the cases $N=2$ (left) and $N=6$ (right).
	The left hand axis where $m = \infty$ corresponds to pure
	Yang-Mills theory, where a ``confining/deconfining''
	phase transition occurs at $L_{\rm c} \sim \Lambda^{-1}$.
	Phases are labeled according to the \emph{unbroken} subgroup of
	$\Z_N$ center symmetry;
	the ``$\Z_1$'' region is the phase with totally broken center
	symmetry.
	The slopes of the transition line(s) approaching the origin are
	perturbatively computable.
	Within the center symmetric phase,
	the limit of vanishing size $L$ (bottom axis)
	is expected to be continuously connected
	to the large $L$, large $m$ domain (upper left corner)
	which corresponds to decompactified Yang-Mills theory.
	The dot on the right-hand boundary indicates the
	chiral symmetry transition point of the massless theory,
	expected to occur  at an $O(1/(N\Lambda))$ value of $L$.
	}
        \label {fig:phase1}
        \end{center}
        }
    }
\end{FIGURE}

With fewer than $\nf^*$ flavors,
the decompactified theory on $\R^4$, for any value of fermion mass,
has a typical confining phase with a strong scale of $\Lambda$.
Consequently, for $L$ large compared to $1/\Lambda$
the center symmetry will be unbroken.
This absence of any center symmetry breaking
for $L \gg 1/\Lambda$ should hold for all values of $m$.

Heavy fermions, with $m \gg \Lambda$, make a negligible
contribution to dynamics on the scale of $\Lambda$,
and so the difference between imposing periodic and antiperiodic
boundary conditions on such fermions is also negligible.
Consequently, for large $m$ there will be 
a conventional confining/deconfining phase transition
at $L = L_{\rm c} = O(1/\Lambda)$,
where $L_{\rm c} = 1/T_{\rm c}$ is the inverse
transition temperature in pure Yang-Mills theory.
Across this transition, the center symmetry changes from completely
broken ($L < L_{\rm c}$) to fully restored
($L > L_{\rm c}$).
The completely broken phase just below $L_{\rm c}$ presumably
connects directly to the completely broken phase at $L > L_1$
identified in the small $L$ analysis.

As noted above, the conventional understanding of confinement
in QCD-like theories (on $\R^4$) implies that
a center-symmetric phase will be present
at vanishingly small $m$ and sufficiently large $L$.
The perturbative analysis valid when $\Lambda \ll 1/(NL)$
shows that a center-symmetric phase is also present when
the fermion mass is sufficiently small, $m \lesssim 1/(NL)$.
The most plausible scenario
is that these center symmetric regions 
are part of a single connected center symmetric phase
which exists for all $L$ when the fermions are sufficiently light,
$m < m^* = O(\Lambda)$.%
\footnote
    {
    The chiral limit of QCD(adj) with periodic boundary conditions
    on $S^3 \times S^1$ was studied in Ref.~\cite{Unsal:2007fb},
    where it was argued that there is no center symmetry changing
    phase transition, consistent with expectations in the
    partially decompactified $\R^3 \times S^1$ limit.
    }
The resulting phase diagram, as a function $L$ and $1/m$,
is illustrated in Fig.~\ref{fig:phase1}
for two representative values of $N$.

\subsection*{Single flavor: \boldmath $\nf = 1$}

Single flavor QCD(adj), in the massless limit,
is $\None$ supersymmetric Yang-Mill theory.
Turning on a small but  non-vanishing mass corresponds to a soft
breaking of supersymmetry.
Exact supersymmetry implies that
the one loop potential vanishes at $m=0$, as one
may easily confirm after substituting $\nf = 1$ into the result
(\ref{eq:pot1}) and sending $m\to 0$.
For small but non-zero fermion mass,
the sub-leading term in the small $z$ asymptotic behavior
(\ref{asymptotes}) contributes
and the $\nf = 1$ effective potential becomes
 \begin{align}
    V[\Omega]&= -
    \frac{2m^2}{\pi^2L^2}
    \sum_{n=1}^{\infty}   \frac{1}{n^2} \,
    \left|\tr\, \Omega^n \right|^2 + O(m^4) \,.
\label{eq:pot4}
\end{align}
Within the domain of validity of the perturbative analysis,
this shows that Wilson lines with all winding numbers are unstable
when the fermion mass is non-zero,
despite the periodic boundary condition for fermions.
Consequently, the $\nf = 1$ theory  at any non-zero mass $m$
and sufficiently small $L$ will have completely broken center symmetry.

When $m = 0$ there is no perturbative contribution (at any order)
to the Wilson line effective potential, but there is a non-perturbatively
induced effective potential which ensures unbroken center symmetry
in the supersymmetric theory on $\R^3 \times S^1$
\cite{Davies:2000nw} .
For small but non-zero $m$, and small $L$, there will be competition
between the one-loop $O(m^2)$ soft supersymmetry breaking potential
and the non-perturbatively induced superpotential,
leading to non-uniformity in the $m\to 0$ and $L \to 0$ limits.
The transition line separating center-symmetric and completely broken
phases must emerge from the $L=m=0$ corner of the phase diagram,
as illustrated for an $SU(2)$ theory on the left side of Fig.~\ref{fig:phase2}.
Unlike the previous multi-flavor case, for larger values of $N$
there is no reason to expect the phase diagram to contain any region
with partially broken center symmetry.

\begin{FIGURE}[ht]
    {
    \parbox[c]{\textwidth}
        {
        \begin{center}
        \includegraphics[scale=0.5]{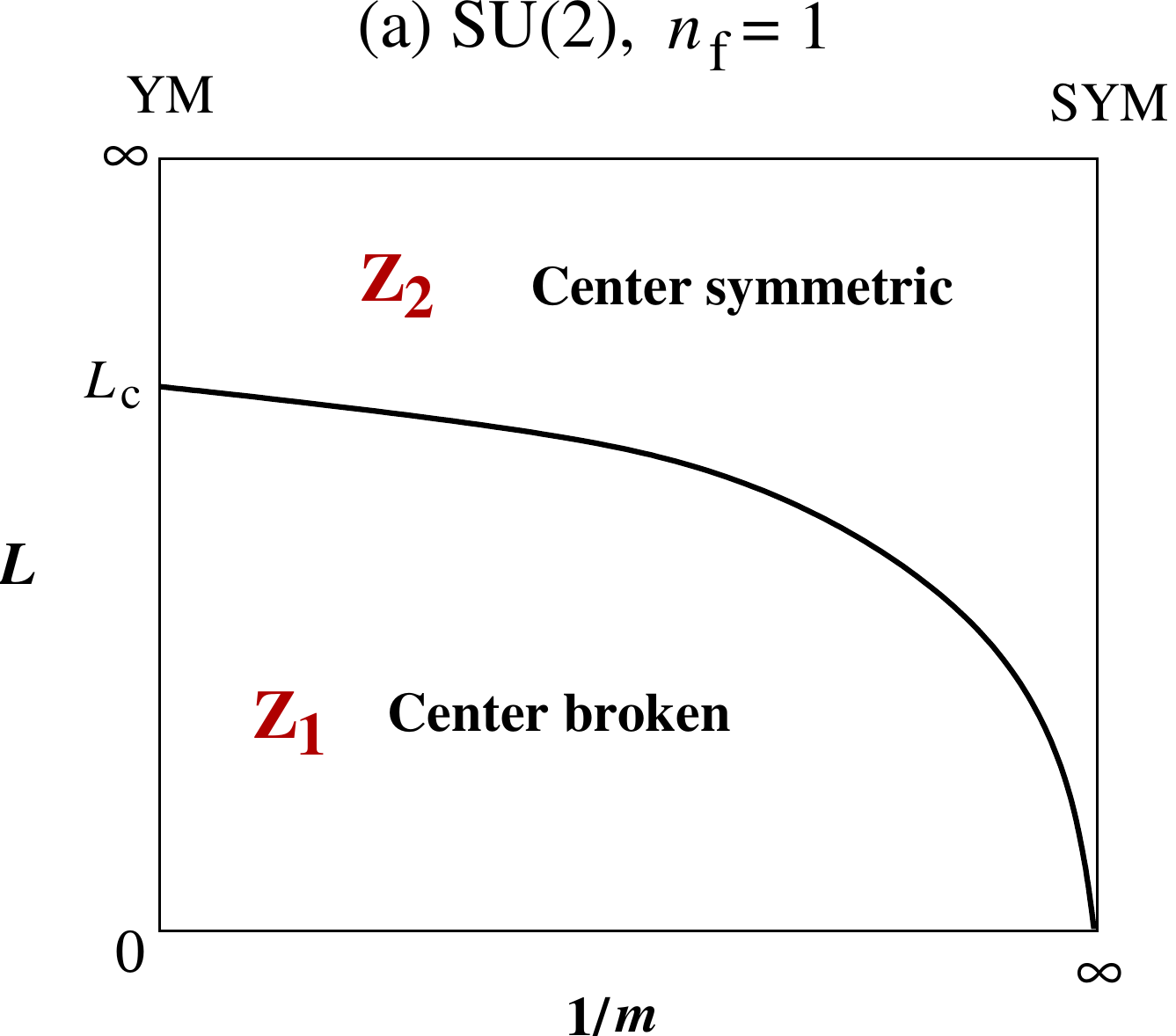}
	\hfil
	\hfil
        \includegraphics[scale=0.5]{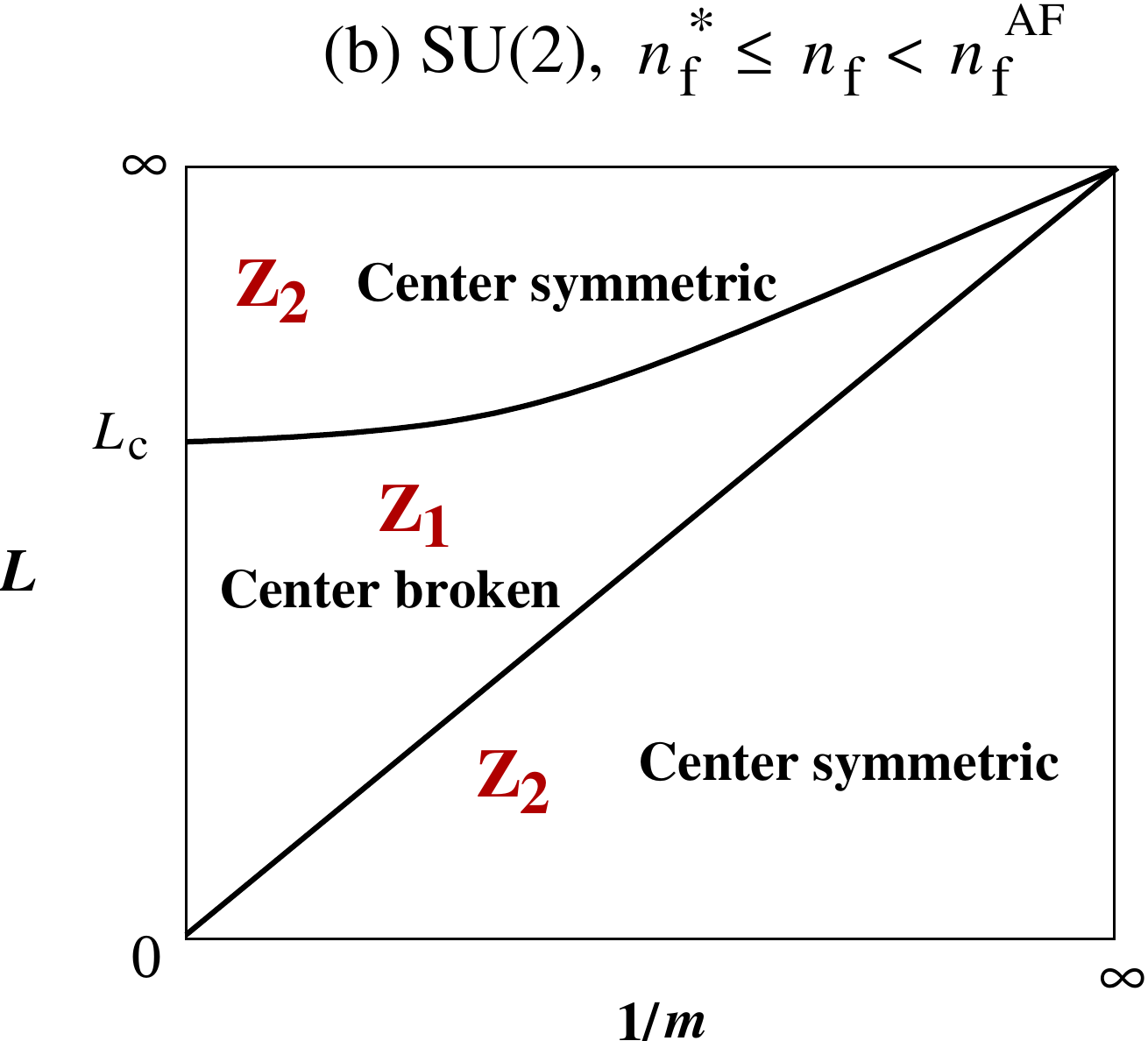}
	\vspace*{-10pt}
        \caption
	{
	Center symmetry realization of QCD(adj)
	on $\R^3 \times S^1$ with periodic spin connection,
	as a function of compactification size $L$
	and inverse fermion mass $1/m$,
	for the case of a single flavor (left),
	or multiple flavors within the conformal window (right),
	$\nf^* \le \nf < \nf^{\rm AF}$.
	For $\nf = 1$, the right-hand $m=0$ axis corresponds to
	$\None$ SYM, which
	does not break center symmetry for any $L$.
	With non-zero fermion mass $m$, a single adjoint
	fermion is insufficient to stabilize the center symmetry at
	arbitrarily small $L$.
	For $\nf^* \le \nf < \nf^{\rm AF}$ (right),
	the phase boundaries separating center symmetry broken
	and unbroken phases are expected to
	extend to $L = \infty$ as $m \to 0$,
	with no continuous connection between
	the center-symmetric phases at large and small $L$.
	}
        \label{fig:phase2}
        \end{center}
        }
    }
\end{FIGURE}

\subsection*{Multiple flavors: \boldmath $\nf^* \le \nf < \nf^{\rm AF}$}

For this range of flavors, the chiral limit of the theory
is in the conformal window, with long distance behavior (on $\R^4$)
described by a non-trivial renormalization group (RG) fixed point.
When the theory is compactified on $\R^3 \times S^1$ and a
non-zero fermion mass is introduced the scale invariant
long distance behavior is cut-off, either by $m$ or $1/L$.
The resulting behavior in the $(L,m)$ phase diagram
will be qualitatively identical to the case $2 \le \nf < \nf^*$
discussed above --- except in the corner where $m \to 0$ and $L \to \infty$.
Specifically, nothing changes in the small $L$ perturbative analysis.
For all finite values of $m$, a center symmetric phase will be present
for sufficiently small compactification size.
One or more transition lines (depending on the value of $N$)
emerge from the $L = 1/m = 0$ corner of the phase diagram
with slopes given by Eq.~(\ref{eq:seq}).
In the large mass, large $L$ portion of the phase diagram,
$m \gg \Lambda \gtrsim 1/L$, the theory reduces to pure Yang-Mills
theory and exhibits a confinement/deconfinement transition
at $L = L_{\rm c} = O(1/\Lambda)$.

In the massless limit, on $\R^4$, the gauge coupling $g^2(\mu)$
ceases to run logarithmically below the scale $\Lambda$
and asymptotes to a non-zero fixed-point value $g^2_*$.
Ref.~\cite{Poppitz:2009uq} argues that the fixed point coupling will
be relatively weak, and that the massless theory on $\R^3 \times S^1$
will have unbroken center symmetry for all $L$.
Turning on a small fermion mass $m \ll \Lambda$
will have negligible effect on the RG flow for $\mu \gg m$,
but will eliminate the fermion contribution to the RG
evolution at scales small compared to $m$.
Since the massless fermion contribution to the beta function
was essential for producing a fixed point, the effective
gauge coupling will only remain (nearly) constant for
scales between $\Lambda$ and $m$.
Below the scale $m$ the coupling will resume its increase,
growing just as it does in pure Yang-Mills theory and eventually
becoming strong and driving confinement at a scale
$\Lambda_{\rm YM} \sim m \, e^{-8\pi^2/(b_0 g_*^2)}$.

When the mass-deformed theory is compactified on $\R^3 \times S^1$,
the confining phase will be stable, and center symmetry unbroken,
provided $1/L \ll \Lambda_{\rm YM}$.
If the prediction of Ref.~\cite{Poppitz:2009uq}
of a rather small value of the fixed point coupling is correct,
then $\Lambda_{\rm YM}$ will be substantially smaller than $m$
(even though there is no arbitrarily large parametric separation).
Assuming so,
a conventional confining/deconfining phase transition,
just as in pure Yang-Mills theory, must occur when $L$ decreases to
$L_{\rm c} = O(1/\Lambda_{\rm YM})$,
since the fermions 
have negligible effect on physics at scales well below $m$.
When the compactification size decreases further to $O(1/m)$ or smaller,
the fermions can influence the resulting symmetry realization.
When the fermions (with non-thermal periodic boundary conditions)
are sufficiently light compared to $1/L$ they will stabilize
the center symmetric phase, just as in the massless limit.

The simplest possibility for the resulting phase diagram
is sketched on the right hand side of Fig.~\ref{fig:phase2},
for the case of $N = 2$.
The novel feature is the ``sliver'' of totally broken center
symmetry phase extending all the way to $L = \infty$ and $m = 0$,
surrounded on either side by a phase with unbroken center symmetry.
For $N > 3$, there will also be phases with partially
broken center symmetry emerging from $L=1/m=0$ and lying
in between the totally broken and small-$L$ symmetric phases,
just as in Fig.~\ref{fig:phase1}(b).
We expect these partially broken phases to terminate at a non-zero
$O(\Lambda)$ value of $m$, but cannot exclude the logical possibility that
they also persist in slivers extending to the upper right $m=1/L=0$ corner.

\subsection*{Large \boldmath $N$ limit of massive QCD(adj)}


As $N\to\infty$, $\Z_N$ center symmetry transformations
become a dense set within the continuous $U(1)$ group.
At $N = \infty$, any discrete cyclic group $\Z_p$ may be
regarded as a subgroup of the center symmetry.
For $\nf > 1$,
when the compactification size $L$ lies in the interval
$[L_p, L_{p-1}]$, then the effective potential
(\ref{eq:pot1}) is minimized when the eigenvalues of $\Omega$ form
$p$ clumps equi-spaced around the unit circle,
so that $\tr\,\Omega^p \ne 0$
while all traces of $\Omega$ to lower powers vanish.
At these minima, the center symmetry is broken to $\Z_p$.
For fixed $m \gg \Lambda$,
as $L$ decreases from $O(1/\Lambda)$ to zero,
passing through each $L_p \propto 1/p$,
there is an infinite sequence
of phase transitions with an accumulation point at $L = 0$:
\begin{align}
    \advance\arraycolsep 2pt
    \begin{array}{lcccccccccccccc}
    \mbox{Compactification size}:
    && L_1
    &>& L_2
    &>& L_3
    &>& \cdots
    &>& L_{p-1}
    &>& L_p
    &>& \cdots
    \\
    \mbox{Residual center symmetry}:
    & \Z_1 & |
    & \Z_2 & |
    & \Z_3 & |
    && \cdots && |
    & \Z_p & |
    && \cdots
    \end{array}
\end{align}
For very large but finite values of $N$,
not all of these intervals will correspond to distinct
symmetry realizations.
Which transitions are associated with changes in symmetry
realization will depend on the prime factorization of $N$.
But within each interval $[L_p,L_{p-1}]$,
the eigenvalues of $\Omega$ will form $p$ clumps ---
some having $\lfloor N/p \rfloor$ eigenvalues and others
having $\lceil N/p \rceil$.
In other words,
the finite $N$ eigenvalue distribution will
approximate $p$ equi-distributed clumps as closely as possible.
The Wilson line expectation value
$\langle\frac 1N\, \tr\, \Omega^k\rangle$
will be $O(1)$ when the winding number $k$ is an integer multiple of $p$,
and $O(1/N)$ when the winding $k$ is not a multiple of $p$.

As discussed above, for arbitrarily large fermion mass,
an unbroken center symmetry phase will exist
both at large compactification size,
$L \gtrsim O(1/\Lambda)$, and sufficiently small size,
$L < L_{\lfloor N/2 \rfloor}$ when $\nf > 1$.
Consequently,
one might hope that large $N$ volume independence would relate
massive QCD(adj) on $\R^4$ to the theory in the small $L$
center symmetric domain.
Since properties of QCD(adj) with $m \gg \Lambda$ differ
negligibly from pure Yang-Mills theory, this suggests that
massive QCD(adj) with sufficiently small $L$ and large $m$
could serve as a reduced model exactly reproducing properties
of large-$N$ Yang-Mills theory.
This, however, is not the case on  $\R^3 \times S^1$. 
The problem is that the condition for unbroken center symmetry
in the reduced theory, $L < L_{N/2} = 2z_*/(Nm)$,
combined with the \emph{strong coupling} condition (\ref{eq:ana})
for large $N$ volume independence,
$N L \Lambda \gg 1$, together imply that $m \ll \Lambda$.
In other words, 
when taking $N \to \infty$ for fixed mass $m \gg \Lambda$,
staying within the center symmetric small-$L$ phase forces
one to send $L$ to zero, $L \lesssim 1/(Nm)$,
and this makes it impossible to remain
within the $NL\Lambda \gg 1$ regime required for
large $N$ volume independence.
This means that massive QCD(adj) on small $S^1 \times \R^3 $. 
cannot be used as a large $N$
reduced model for (continuum) Yang-Mills theory on $\R^4$.

Nevertheless, provided the phase diagram sketched in
Fig.~\ref{fig:phase1} is
qualitatively correct as $N \to \infty$,
large $N$ volume independence is valid in massive QCD(adj)
with $2 \le \nf < \nf^*$,
for all $L$,
when the fermion mass is below the $O(\Lambda)$ lower limit $m^*$
of any broken center symmetry phase.
And likewise for $\nf^* \le \nf < \nf^{\rm AF}$ and $m < O(\Lambda)$,
large $N$ volume independence will relate the center-symmetric phases
at large and small $L$,
despite the presence of an intervening broken symmetry phase.

\section {Multiple compactified dimensions}

We now turn to a consideration of QCD(adj)
on $\R^2 \times (S^1)^2$.
The extension to compactifications on higher dimensional torii
will be discussed below.

Let $\Omega_1$ and  $\Omega_2$  denote the holonomy of the gauge field
around the two independent cycles of the 2-torus $T^2 = (S^1)^2$.
The set of vacua of the classical gauge theory is the space of flat
connections,
$F_{\mu \nu}=0$,
which implies commuting covariantly constant holonomies,
\begin{equation}
    [ \Omega_1, \Omega_2]=0 \,.
\end{equation}
Hence, for vacuum configurations
the two holonomies can be simultaneously diagonalized by a suitable
gauge transformation.
In other words,
the classical vacua may be parametrized by two sets of $N$ eigenvalues,
\begin{subequations}
\begin{align}
    \Omega_1 &= 
    {\rm diag}
    \left( e^{i\alpha_1},\, e^{i\alpha_2},\, \cdots,\, e^{i\alpha_N} \right) ,
\\
    \Omega_2 &= 
    {\rm diag}
    \left( e^{i\beta_1},\, e^{i\beta_2},\, \cdots,\, e^{i\beta_N} \right) ,
\end{align}%
\label{backg}%
\end{subequations}
up to a global gauge transformation
(which simultaneously conjugates $\Omega_1$ and $\Omega_2$).
Each diagonalized holonomy $\Omega_i$ may be regarded as taking
values in the maximal torus of $SU(N)$, which we label as ${\bf T}_N$.
For generic configurations with distinct eigenvalues,
the subgroup of global gauge transformations which preserve 
the diagonalized form (\ref{backg})
is the Weyl group $\mathcal W$ of $SU(N)$, whose elements
\emph{simultaneously} permute the eigenvalues of the two holonomies,
$(\alpha_i, \beta_i) \rightarrow (\alpha_{\sigma(i)}, \beta_{\sigma(i)}) $,
where $\sigma \in \SN$ (with $\SN$ the $N$-element permutation group).
Consequently,
the classical moduli space of the theory can be described as
\begin{equation}
{\cal M} = \left( {\bf T}_N \right)^2 / \SN \,.
\label{eq:T2moduli}
\end{equation}
Another useful way to describe  ${\cal M}$ is as follows.
Let $\widetilde T^2$
denote the torus dual to the base space 2-torus.
If $\vec L \equiv (L_1,L_2)$ are the periods of $T^2$, then
the dual torus $\widetilde T^2$ has periods
$\widetilde L_1 = {2\pi}/{L_1}$ and
$\widetilde L_2 = {2\pi}/{L_2}$.
Each pair of simultaneous eigenvalues $(e^{i\alpha},\, e^{i\beta})$
may be written as $(e^{i k^1 L_1},\, e^{i k^2 L_2})$
where $\vec k \equiv (k^1,\, k^2) \in \widetilde T^2$.
Consequently,
the moduli space ${\cal M}$ may also be viewed as the configuration
space of $N$ identical ``eigenparticles" which are placed on the
dual torus $\widetilde T^2$.
(Clearly, this description generalizes immediately to compactifications
on higher dimensional torii.)

Integrating out fluctuations will produce an effective potential
which depends on the commuting holonomies $\Omega_i$.
In QCD(adj) with $\nf$ fermions (having a common mass $m$ and periodic
spin connection) the one-loop result may be expressed as
a sum of pairwise interactions between eigenparticles,
\begin{equation}
    V[\Omega_1,\Omega_2]
    =
    \half
    \sum_{i,j=1}^N \>
    v [\vec\alpha_i {-} \vec \alpha_j] \,,
\label{eq:Veff}
\end{equation}
with
\begin{equation}
    v[\Delta\vec \alpha]
    \equiv
    {\sum_{\vec n \in \Z^2}}' \>
    m_{\vec n}^2 \, \cos[\vec n \cdot \Delta\vec\alpha] \,,
\label{eq:T2vij}
\end{equation}
and
\begin{equation}
    m_{\vec n}^2
    \equiv
    \frac 2{\pi^2 |\vec {Ln}|^4}
    \left[
	-1 + \half \nf \, (m |\vec {Ln}|)^2 K_2 (m |\vec {Ln}|)
    \right] .
\label{eq:m2T2}
\end{equation}
Here
$\vec\alpha_i \equiv (\alpha_i,\beta_i)$,
$\vec n = (n_1, n_2)$,
$\vec {Ln} \equiv (L_1 n_1, L_2 n_2)$,
and the prime on the sum indicates omission of
the term with $\vec n = 0$.
Alternatively, one may express this potential as a sum of
squares of Wilson lines,
\begin{equation}
    V[\Omega_1,\Omega_2]
    =
    \half
    {\sum_{\vec n \in \Z^2}}' \>
    m_{\vec n}^2 \,
    \big| \tr (\Omega_1^{n_1} \Omega_2^{n_2} ) \big|^2 \,.
\label{eq:VeffT2}
\end{equation}
Unbroken $(\Z_N)^2$ center symmetry implies
vanishing of the traces $\tr(\Omega_1^{n_1} \Omega_2^{n_2})$
for all winding numbers $n_1$ and $n_2$ which are not multiples of $N$.
The fermion contribution (proportional to $\nf$)
to the effective mass squared $m_{\vec n}^2$ is positive.
{}One may see from the pairwise potential (\ref{eq:T2vij})
that this means fermions generate repulsive interactions
between every pair of eigenparticles.
The equivalent form (\ref{eq:VeffT2}) shows that
the positive fermion contribution to $m_{\vec n}^2$
works to stabilize the center symmetric configuration
with vanishing traces.
In contrast, 
the ($-1$) pure gauge contribution to $m_{\vec n}^2$
produces attractive interactions
between eigenparticles which
work to destabilize center symmetry.
This exactly parallels the situation on $\R^3 \times S^1$,
except that eigenvalue pairs on $\widetilde T^2$ take the
place of single eigenvalues on $S^1$.

Naturally,
the competition between fermion and gauge field contributions
leads to different types of minima depending on the values of
the periods $\vec L$, the fermion mass $m$, and
the number of flavors $\nf$.
For sufficiently large fermion mass
(within the small $L$ domain of validity of the perturbative analysis),
the gauge field contribution will dominate,
causing all $N$ eigenparticles to coalesce
and producing complete spontaneous breaking of the $(\Z_N)^2$ center symmetry.
For sufficiently light and numerous fermions,
the repulsive fermion-induced interaction between eigenparticles
will dominate.
This will cause the eigenparticles to disperse throughout
the dual torus.
Finding
the arrangement which minimizes the effective potential
is analogous to a sphere-packing problem;
the precise result will depend sensitively on $N$
and the aspect ratio $L_1/L_2$.
When, for example, $L_1 = L_2 \equiv L$ and $N$ is a perfect square,
the minimum of $V[\Omega_1,\Omega_2]$ for sufficiently light
mass and $\nf > 1$ should correspond to ``crystallization''
of the $N$ eigenparticles into a $\sqrt N \times \sqrt N$
square array.%
\footnote
    {
    More explicitly, this means
    $
    \Omega_1 \propto C_{\sqrt N} \otimes 1_{\sqrt N}
    $
    and
    $
    \Omega_2 \propto 1_{\sqrt N}  \otimes  C_{\sqrt N}
    $,
    where
    $1_{\sqrt N}$ is a $\sqrt N$-dimensional identity matrix and
    $
	C_{\sqrt N} =  \mathrm{diag}\,
	( 1,\, \omega,\, \omega^2,\, \cdots,\,  \omega^{{\sqrt N} -1} )
    $,
    with $\omega \equiv \exp(2\pi i/\sqrt N)$
    a $\Z_{\sqrt N}$ generator.
    However,
    we have not checked to see if, for example,
    a bcc lattice is preferred over a simple cubic lattice.
    For suitable values of $N$ the minimum energy configuation
    of eigenvalues will be some regular lattice; exactly which
    lattice is inessential to our discussion.
    }
Such a configuration produces vanishing order parameters,
$\tr\,(\Omega_1^{n_1} \Omega_2^{n_2}) = 0$,
for windings $n_1$ and $n_2$ which are non-zero modulo $\sqrt N$,
corresponding to spontaneous breaking of the
$(\Z_N)^2$ center symmetry down to a $(\Z_{\sqrt N})^2$ subgroup.
For these configurations, the evaluation of the effective potential
is reliable provided $\sqrt N L \Lambda \ll 1$;
this is the condition that the coupling be weak on the
$O[1/(\sqrt N L)]$ scale of the lightest off-diagonal degrees of freedom
(\emph{i.e.}, charged $W$'s) which were integrated out to
obtain the effective potential.

More generally, for $O(1)$ values of $L_1/L_2$
and large values of $N$,
the minimum of the effective potential will correspond to
configurations where the eigenparticles are
distributed with approximately uniform density on the dual torus.
The typical nearest-neighbor separation between eigenparticles
on the dual torus will be $\sim 2 \pi /\sqrt N L$,
where $L$ is the geometric mean of $L_1$ and $L_2$,
and the appropriate weak coupling condition will again
be $\sqrt N L \Lambda \ll 1$.

For a highly anisotropic torus,
$L_2 \gg L_1$, the eigenparticles will form
a one-dimensional array.
When, for example, $L_2 = p L_1$ for some large integer $p$ and
$M \equiv N/p$ is an integer factor of $N$, then
$\vec \alpha_j = (2\pi j/N,2\pi j/M)$ is a minimum energy configuration.
This configuration is
invariant (up to Weyl permutations) under a $\Z_N$ subgroup obliquely
embedded within the $\Z_N \times \Z_N$ center symmetry group and
corresponds to a partial breaking of the $\Z_N \times \Z_N$ symmetry
down to a single $\Z_N$.

In contrast to the situation on $\R^3 \times S^1$,
no configuration of eigenparticles on the dual torus can
be invariant under the entire $(\Z_N)^2$ center symmetry.
The unbroken subgroup of the center symmetry will be some
$\Z_p \times \Z_q$ with $p q \le N$.
Consequently, there will be (at least) $O(N)$ gauge-inequivalent
degenerate minima, related by center transformations.
But before one can conclude that this
truly implies spontaneous breaking of center symmetry,
the effects of fluctuations must be considered.
On $\R^2 \times T^2$, fluctuations of eigenvalues away from the
minimum energy configurations behave like two-dimensional scalar
fields.
These fluctuations, for $\nf > 1$,
acquire a non-zero mass of order $\sqrt\lambda/L$
from the one-loop effective potential,
with $\lambda \equiv g^2 N$ the 't Hooft coupling.
When $L$ is small and the perturbative evaluation of the effective potential
is reliable,
the probability of such fluctuations overcoming the action
barriers separating different minima is negligible.
Consequently, on $\R^2 \times T^2$, 
multiple light adjoint representation fermions cannot
prevent at least partial breaking of center symmetry
in the limit of small compactification radii.
A rich pattern of phase boundaries will emerge from
the $L_1 = L_2 = 1/m = 0$ corner of the phase diagram,
just as for the $\R^3 \times S^1$ case, when $\nf > 1$.

Once again, $\nf = 1$, or single flavor QCD(adj),
is a special case.
With a non-zero mass $m$, the center-symmetry stabilizing
fermion contribution can never dominate the destabilizing gauge
field contribution in the coefficients (\ref{eq:m2T2}).
Consequently, as the compactification sizes $L_1, L_2 \to 0$
for fixed mass $m > 0$, the center symmetry will be completely broken,
just like the situation on $\R^3 \times S^1$.

In the massless limit, the $\nf = 1$ theory becomes
$\None$ supersymmetric Yang-Mills theory.
Center symmetry is surely unbroken when the
compactification size is sufficiently large.
In supersymmetric theories with
supersymmetry preserving boundary conditions,
it is believed that
there are no phase transitions as a function of volume.
The singularities in the holomorphic coupling are of (real) co-dimension two
\cite{Seiberg:1994aj,Intriligator:1994sm},
so even if a singularity was encountered
as the compactification size decreased, 
it is possible to analytically continue around it.
Hence, center symmetry in $\None$ SYM
(compactified on $\R^2 \times T^2$)
should be unbroken for any value of $L_1$ and $L_2$.

It is instructive to see how this conclusion can emerge
from a small $L$ analysis.
The perturbative effective potential vanishes identically
in this supersymmetric theory but, as noted above,
a non-perturbative superpotential is generated on $\R^3 \times S^1$
\cite{Davies:2000nw}, and hence necessarily also on $\R^2 \times T^2$.
(This follows from considering the regime $L_1 \ll L_2$,
where physics must approach the $\R^3 \times S^1$ case.)
The parametric size of the resulting bosonic potential
is $\Lambda^6 L^2$ where $L = \min(L_1,L_2)$.
This potential lifts the classical moduli space
(\ref{eq:T2moduli}) but, for $\Lambda L \ll 1$,
it is a very weak ``pinning'' potential for the eigenparticles
and generates an effective mass $\mu$ for fluctuations of eigenvalues
which is tiny,
$\mu \sim O(\sqrt\lambda \Lambda^3 L^2)$.
The resulting fluctuations, on scales which are large compared to
$L$ but small compared to $1/\mu$,
have amplitudes large enough to wash out the distinction
between neighboring degenerate minima related by center transformations.%
\footnote
    {%
    Consider, for example, the specific case of $N = M^2$ and $L_1 = L_2 = L$
    where the minimum energy configuration of eigenparticles should be
    a square lattice.
    Eigenvalue fluctuations behave like massless $2d$ fields,
    with
    $
	\langle \delta \alpha(x) \delta \alpha(0) \rangle
	\sim
	(\lambda/4\pi N) \ln 1/(\mu^2 x^2)
    $,
    for $|x| \lesssim 1/\mu$.
    Examining, \emph{e.g.}, the order parameter
    $\mathcal O \equiv \frac 1N \tr (\Omega_1^M)$,
    the difference in $\langle \mathcal O \rangle$
    between neighboring minima is $|e^{2\pi i/M} - 1| = O(1/M)$.
    If $\mathcal O_d$ denotes $\mathcal O$ averaged over a region
    of size~$d$, one finds
    $
	\langle (\delta \mathcal O_d)^2 \rangle
	\sim
	\frac 1N
	[\frac \lambda {4\pi} \ln \frac 1{\mu^2 d^2}]^2
    $.
    So the rms fluctuation in $\mathcal O_d$ is $O(1/M)$ or larger when
    $d \le \bar d \sim \mu^{-1} e^{-2\pi/\lambda}$.
    Inserting
    $
	\mu^{-1}
	\sim (L/\sqrt \lambda) (\Lambda L)^{-3}
	\sim (L/\sqrt \lambda) \, e^{8\pi^2/\lambda}
    $
    shows that
    $
	\bar d \sim (L/\sqrt\lambda) \,
	e^{(8\pi^2 - 2\pi)/\lambda}
    $,
    where $\lambda = \lambda(1/L)$.
    So for small $L$, there is a parametrically large
    hierarchy, $L \ll \bar d \ll 1/\mu$.
    }
Hence, for $\None$ SYM, a classical analysis of the
Wilson line effective potential is invalid;
the effects of nearly massless two-dimensional eigenvalue
fluctuations cannot be neglected.
Integrating out the effects of these fluctuations on length
scales between $L$ and $\mu^{-1}$ will modify the effective potential
relevant for even longer scales, leading to merging
of the multiple minima and restoration of the full center symmetry.

Returning to the case of $\nf > 1$,
we noted above that for arbitrarily small $L$
there will be at least partial breaking of the center symmetry.
The implications of this for large $N$ volume independence is,
perhaps surprisingly, negligible.
With multiple fermion flavors and sufficiently small compactification size,
all coefficients $m_{\vec n}^2$
in the Wilson line effective potential (\ref{eq:VeffT2})
with up to $N$ windings are positive,
favoring unbroken center symmetry.
But no configuration of eigenvalues, for finite $N$, can force all
order parameters $\tr (\Omega_1^{n_1} \Omega_2^{n_2})$ to vanish
(for windings $n_1$ and $n_2$ non-zero modulo $N$).
Nevertheless when,
for example, $N = M^2$ and $L_1 = L_2$, the unbroken
$\Z_{\sqrt N} \times \Z_{\sqrt N}$ subgroup of center symmetry
still forces all loops with winding numbers less than $\sqrt N$
to vanish.
As $N \to \infty$, this (plus unbroken translation symmetry)
is sufficient to guarantee that the loop equations of the
compactified and decompactified theories coincide.
For other values of $N$ or $L_1/L_2$,
the minimum energy configurations of eigenparticles on the dual torus
$\widetilde T^2$ will lead to non-zero values of Wilson lines
even when the winding numbers are small.
But repulsion of eigenparticles on the dual torus
will cause the 
order parameters $\frac 1N \tr (\Omega_1^{n_1} \Omega_2^{n_2})$
to be $O(1/N)$ --- not $O(1)$ --- provided $n_1, n_2 \ll N$.
This means that the large $N$ limit of the expectation value
of any topologically non-trivial Wilson loop vanishes,
showing that the full center symmetry is effectively restored
as $N = \infty$.
Consequently, large $N$ volume independence will relate $\nf \ge 2$
QCD(adj) on $\R^2 \times T^2$ with sufficiently small compactification
size to the decompactified theory on $\R^4$.

Most aspects of the above discussion of QCD(adj) on $\R^2 \times T^2$
generalize immediately to the case of $\R \times T^3$ or $T^4$.
The only difference in expressions (\ref{eq:Veff})--(\ref{eq:VeffT2})
is that $\vec \alpha$ and $\vec n$ change from two-component
to three- or four-component vectors.%
\footnote
    {
    There is one noteworthy change
    for $T^4$ compactifications:
    the sum over $\vec n \in \Z^4$ which now appears
    in the coefficients (\ref{eq:T2vij})
    is only conditionally convergent for $\Delta\vec\alpha \ne 0$,
    and has a logarithmic divergence at $\Delta\vec\alpha = 0$.
    This reflects the fact that coinciding eigenparticle positions
    on the dual torus represent special configurations where
    some of the off-diagonal fluctuations which were integrated
    out to produce the effective potential (\ref{eq:Veff}) become massless.
    On $T^4$ with coinciding eigenvalues,
    these are off-diagonal zero-modes for whom quartic interactions
    remain relevant.
    Inappropriately applying a Gaussian approximation to these zero modes
    leads to the unphysical logarithmic singularity in the effective
    potential.
    }
However,
fluctuations of eigenvalues become progressively more important
as the number of uncompactified dimensions decreases.
For finite values of $N$, the discrete center symmetry cannot break
spontaneously when the theory is compactified on $\R \times T^3$ or $T^4$.
However, as is well known, spontaneous symmetry breaking can occur
at $N = \infty$ even in finite volume theories.
Consequently, one must carefully examine the effects of
eigenvalue fluctuations as $N \to \infty$.
When repulsive interactions dominate and eigenparticles are
distributed throughout the dual torus, one may show
from the expression (\ref{eq:VeffT2})
that the potential barrier
which separates degenerate minima related by
a center transformation is $O(N^0)$,
and does not grow as $N$ becomes large.
On $T^4$, where there is no infinite volume of
uncompactified dimensions,
this implies that fluctuations
sampling all center-symmetry related minima
will have non-vanishing probabilities as $N \to \infty$,
thereby ensuring unbroken center symmetry at $N = \infty$.
The situation is different when attractive interactions
between eigenparticles dominate and the eigenparticles clump.
In this case, the potential barrier between different minima
is $O(N)$, and the probability of symmetry-restoring
fluctuations vanishes at $N = \infty$.

On $\R \times T^3$, the relevant fluctuations are tunneling transitions
which are localized in the uncompactified dimension,
but the basic conclusion in the same:
when the repulsive fermion-induced interaction between eigenparticles
dominates the attractive gauge field contribution,
fluctuations sampling all center-symmetry related configurations
should retain non-zero probabilities as $N \to \infty$,
so that center symmetry remains unbroken at $N = \infty$.

The basic interaction between eigenparticles
switches between attractive and repulsive at an $O(1)$ value
of $mL$ (for $\nf > 1$).
This is consistent with the numerical work of Bringoltz and Sharpe
\cite{Bringoltz:2009kb}%
\footnote
    {
    For related work, also see Ref.~\cite{Hietanen:2009ex, CGU}.
    }
who investigated a single-site model
of QCD(adj) and found that the center symmetry is intact for
a wide range of fermion mass,
up to an $O(1)$ value of $ma$ (with $a$ the lattice spacing),
for reasonably large values of $N$ and values of the bare 't Hooft
coupling in the range typically used in lattice QCD studies.
For compactifications on $\R \times T^3$ or $T^4$
the bottom line, once again, is that large $N$ volume independence
will be valid when the fermions are sufficiently light and numerous.
But, due to enhanced fluctuations in lower dimensions,
the upper limit on the range of fermion masses for which
unbroken center symmetry persists down to $L \to 0$,
at infinite $N$,
should be  the larger of ($O(1/L)$,  $O(\Lambda)$)
and not  just $O(\Lambda)$ as in 
the earlier cases of $\R^2 \times T^2$ and $\R^3 \times S^1$.
See Refs.~\cite{Azeyanagi:2010ne} for more discussion of this issue.

\section{Prospects}

Volume independence is an exact property of certain large-$N$
gauge theories.
Although the idea is old, it was widely believed that, for  four dimensional gauge theories,  only a
partial reduction was possible
down to a minimal compactification
size $L = L_{\rm c} \sim 1/\Lambda$,
and it has often been asserted that the small volume theory
is weakly coupled whenever $L \Lambda \ll 1$.
This proves not to be the case.
The understanding that emerges from valid examples of volume
independence (as $L \to 0$) leads to a modified picture.
When center symmetry is unbroken,
a weak coupling description is only possible
when $L \Lambda$ is small compared to a positive power of $1/N$.
It is possible for a QCD-like large $N$ gauge theory,
formulated in a box much smaller than the inverse strong scale,
to reproduce infinite volume results.

There is an ongoing effort in the lattice gauge theory community to determine
the lower boundary of the conformal window for various QCD-like theories.
One of the technical issues in all lattice gauge theory simulations is
controlling finite-volume effects and performing reliable infinite volume
extrapolations.
The $1/N$ suppression of finite-volume effects in large-$N$
center symmetric theories allows one to trade a large volume
extrapolation for a large $N$ extrapolation, and should be
helpful for studies of conformal windows in large $N$ theories.

In a lattice formulation of gauge theories, large-$N$ volume independence
implies a non-perturbative equivalence between four-dimensional field
theories and zero dimensional matrix models or one dimensional quantum
mechanics of large-$N$ matrices
\cite{Kovtun:2007py,Unsal:2008ch,Bringoltz:2009kb}.
It may be possible to develop useful techniques 
directly in zero or one dimension
(which are not applicable to higher dimensional field theories)
in order to gain insight into four-dimensional gauge dynamics.
We believe that these applications of large-$N$
volume independence, and undoubtedly others,
merit further consideration.

\acknowledgments
We are indebted to Erich Poppitz and Prem Kumar for valuable discussions. 
M.\"U's work is supported by the
U.S.\ Department of Energy Grant DE-AC02-76SF00515. 
L.G.Y's work is supported, in part, by the
U.S.\ Department of Energy Grant DE-FG02-96ER40956.

\end{document}